# Strategies for controlling through-space charge transport in metal-organic frameworks via structural modifications

*Christian Winkler and Egbert Zojer**

Institute of Solid State Physics, NAWI Graz, Graz University of Technology, Petersgasse 16, 8010 Graz, Austria

In recent years, charge transport in metal-organic frameworks (MOFs) has shifted into the focus of scientific research. In this context, systems with efficient through-space charge transport pathways resulting from π-stacked conjugated linkers are of particular interest. In the current manuscript, we use density functional theory based simulations to provide a detailed understanding of such MOFs, which in the present case are derived from the prototypical $Zn_2$(TTFTB) system. In particular we show that factors like the relative arrangement of neighboring linkers and the details of the structural conformations of the individual building blocks have a profound impact on band widths and charge transfer. Considering the helical stacking of individual tetrathiafulvalene (TTF) molecules around a screw axis as the dominant symmetry element in $Zn_2$(TTFTB)-derived materials, the focus here is primarily on the impact of the relative rotation of neighboring molecules. Not unexpectedly, also changing the stacking distance in the helix plays a distinct role, especially for structures, which display large electronic couplings to start with. The presented results provide guidelines for achieving structures with improved electronic couplings. It is, however, also shown that structural defects (especially missing linkers) provide major obstacles to charge transport in the studied, essentially one-dimensional systems. This suggests that especially the sample quality is a decisive factor for ensuring efficient through-space charge transport in MOFs comprising stacked π-systems.

## 1. Introduction

Metal-organic frameworks (MOFs) are porous, highly crystalline solids consisting of inorganic nodes connected by organic linkers.[1–3] They have been investigated extensively for various applications in fields like gas storage,[4–6] catalysis,[7–9], and gas separation.[10,11] Until recently, comparably little attention has been paid to the electronic properties of MOFs,[12] although electrically conductive MOFs can be relevant as active materials for several applications like electrocatalysis,[13–17] chemiresistive sensing,[18–23] and energy storage.[24–26] Therefore, in recent years the interest in controlling and modifying the electronic properties of MOFs has gained considerable attention.[12,27,28]

On more fundamental grounds, the electronic (and also the optical) properties of a solid are determined (in a first approximation) by its electronic band structure, where MOFs usually show rather flat bands.[27] This is a consequence of the commonly observed weak hybridization between states localized on the organic linkers and states localized on the inorganic secondary building units. A second reason is the small overlap of the π-systems of neighboring organic linkers. This assessment already comprises two strategies for changing the electronic properties of MOFs: one can either focus on improving the bonding between the metal and the ligands (*through-bond coupling*), or one can try to improve the overlap of the π-systems in neighboring linkers (*through-space coupling*).[12,27,29] In the present contribution, we will focus on the latter approach, where a large overlap of neighboring π-electron systems can result in the formation of bands displaying a large dispersion and generating *through-space* charge transport pathways.

The impact of the packing motif of organic π-systems on intermolecular electronic couplings (expressed via transfer integrals and band dispersions) has been thoroughly studied for organic semiconductors, OSCs. Crucial factors identified for these materials are the stacking distance of neighboring molecules (i.e., neighboring π-systems) and the arrangement of neighboring molecules in terms of relative displacements and rotations.[30–37] All these structural changes lead to changes in the orbital overlap between consecutive molecules, which in turn changes the intermolecular electronic couplings. For example, considering dimers of acenes (or quinacridone) and shifting the molecules relative to each other along their long molecular axis one finds that the transfer integrals oscillate as a function of the displacement.[30–34,36,38] The influence of relative rotations of molecular dimers has been investigated primarily in the context of liquid crystals. There it has been found that the transfer integral varies as a function of the rotation angle.[35,37,39–42] Importantly, independent of whether neighboring molecules are shifted or rotated relative to each other, what counts for the intermolecular electronic coupling (transfer integral) is the overlap of the frontier orbitals. This overlap is determined by the orbitals' shape and symmetry. In this context, it has been suggested that when organic semiconductors can assemble without pronounced sterical constraints, exchange repulsion acts as an intrinsic driving force favoring molecular arrangements with particularly small electronic couplings.[32,33] Therefore, developing design strategies "extrinsically" enforcing favorable molecular arrangements have shifted into the focus of current research.[32,33,43–46] Here, MOFs are of particular appeal, as the framework structure offers an additional level of control over the stacking sequence of neighboring molecules, which goes far beyond what is typically achievable in organic semiconductors.

Materialswise, a particularly promising type of MOFs showing *through-space* charge-transport pathways are frameworks consisting of ligands based on tetrathiafulvalene (TTF).[12] Especially for a subgroup of these MOFs, in which the TTF units form helical stacks with comparably close π-stacking, one observes relatively large electrical conductivities.[12,47] For such systems it has also been shown that reducing the S…S stacking distance within the TTF stacks results in significantly increased conductivities.[48–50]

In this work, we will apply dispersion-corrected density functional theory (DFT) calculations to show how the electronic coupling in such TTF based MOFs can be controlled by additional structural parameters, like the relative rotation or slip of neighboring TTF units. The goal of these calculations is to identify stacking motives that maximize *through-space* charge-carrier mobilities. Moreover, we will address the impact of defects like missing linkers, pair formation, and shifted molecules.

### 1.1 Systems of interest

The starting point for this study is the stacking of the TTF cores of $Zn_2$(TTFTB) [47] shown in Figure 1. The linkers ($H_4$TTFTB = tetrathiafulvalene tetrabenzoate) and the metal nodes (forming $Zn_2$(TTFTB)) crystallize in the $P6_5$ space group with a hexagonal unit cell (a=b=19.293 Å and c=20.838 Å). This results in helical TTF stacks (6 TTF molecules per unit cell), where neighboring TTF molecules are rotated by 60° relative to each other and translated by 3.473 Å in stacking direction. Notably, the $6_5$ screw axis is offset from the central ethylene unit of the TTF cores (see the top view of the model system in Fig. 1). This induces an additional shift of neighboring molecules relative to each other perpendicular to the screw axis.[47] Thus, the centers of the TTF molecules are arranged on a helix, whose projection onto the x,y-plane (the plane perpendicular to the stacking direction, z) becomes a circle with a radius r of ~1.6 Å. Consequently, the shifts of neighboring molecules in the x,y plane (Δx and Δy) fulfil the condition that $\Delta x^2 + \Delta y^2 = r^2$. This stacking motif of the TTF cores is determined by the arrangement of the Zn nodes and the carboxylic acid groups. The MOF structure discussed in the main manuscript contains neither solvent molecules nor water molecules (i.e., the MOF is desolvated and dehydrated). For comparison also the MOF with water molecules coordinating to the Zn atoms has been calculated (see Supplementary Material). Comparing this system to the one without water one finds that the helical stacking motif of the TTF molecules essentially remains the same with only minor changes in the atomic coordinates. Consequently, it is not surprising that also the electronic structure of the TTF stack is hardly affected by the presence of the water molecules (see Supplementary Material).

For analyzing the impact of structural changes on the electronic properties of the TTF stacks, we construct several model systems:

First, based on the relaxed structure of desolvated and dehydrated $Zn_2$(TTFTB), we construct a helical model TTF stack by removing all atoms within the MOF structure apart from the TTF cores. The latter were then saturated by attaching H atoms. Subsequently, the positions of these H atoms were relaxed while keeping all other atomic positions fixed. This results in TTF molecules exhibiting the same geometry and stacking motif as in the MOF. In a similar way also the TTFTB model stack has been constructed: All atoms apart from the TTF core and the connected phenylene rings were removed. In essence, this means that the carboxyl groups have been removed from the linker in the MOF structure and replaced by H atoms. The positions of these H atoms then were relaxed in a subsequent step. The relaxed geometry is shown in the Supplementary Material.

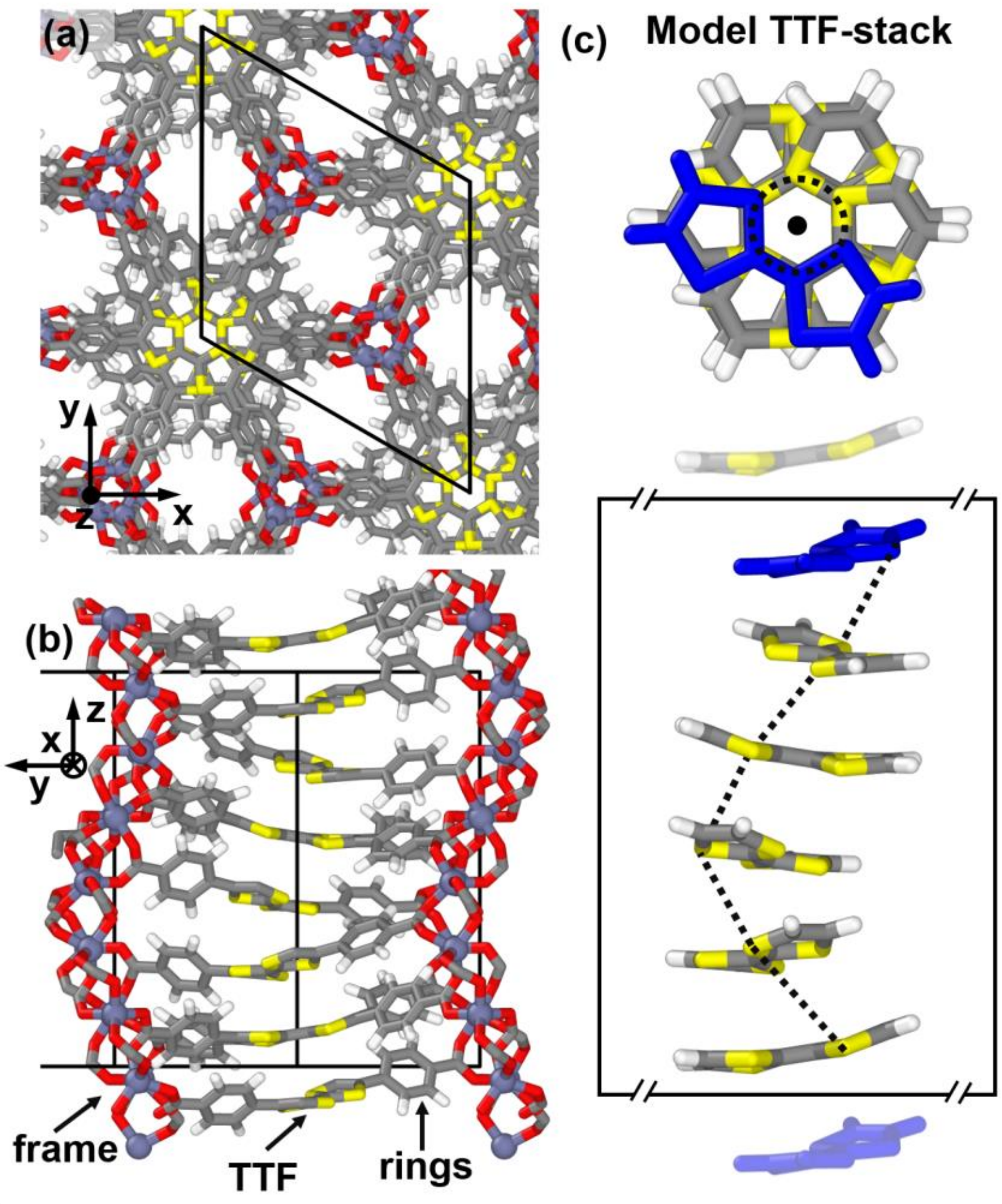

***Figure 1.*** *Structure of $Zn_2$(TTFTB) and the constructed TTF model system. Panels (a) and (b) show the structure of the $Zn_2$(TTFTB) MOF (top and side view). The unit cell of the MOF is represented by thick black lines. Panel (c) contains top and side views of the model system used for describing the one-dimensional charge transport in these materials. The top TTF molecule in the model system is marked in blue, the rotation axis is indicated by the black dot in the center of the panel (c). The dashed black line in the model system indicates the rotation of the molecules around the screw axis. The periodic boundary conditions are indicated by the frame around the repeat unit of the TTF model system. C … grey, H … white, S … yellow, Zn …purple, O … red*

Second, for getting TTF stacks with different numbers of repeat units N, we simply picked one of the symmetry equivalent molecules of the parent stack discussed in the previous paragraph. This molecule was then duplicated, and rotated by the respective rotation angle ((n/N)*360°) around the off-center rotation axis (with n being an integer between 1 and N-1). Then the molecule was shifted in z-direction by n*3.473 Å. This procedure was repeated N-1 times to generate the unit cell for the simulations. For example, for 3 repeat units (N=3) we have in total three TTF molecules per unit cell including the molecule at the bottom of the cell (n=0) and its replicas at n=1 and n=2. The lateral extent of the unit cell of the model system is the same as that observed experimentally for $Zn_2$(TTFTB), while the extent of the unit cell in stacking direction amounts to N*3.473 Å.To verify the construction procedure we compared the electronic structure of the N=6 TTF model stack to the system extracted directly from the MOF structure. The identified differences are almost negligible (see Table 1) and can be assigned to subtle changes in the geometry of individual TTF molecules within the MOF-extracted system. Additionally, molecular dimers were designed in analogy to the construction described in the previous paragraph, but stopping at n=1. As these dimers were simulated using open boundary conditions, any value could be chosen for the rotation angle around the off-center rotation axis. This allowed us to calculate smooth evolutions of the dimer electronic couplings as a function of the relative rotation of neighboring molecules.

Two additional MOFs are also considered in this work: One system is $Cd_2$(TTFTB), which is reported to be isostructural to $Zn_2$(TTFTB), but shows a higher electrical conductivity.[49] The second additional MOF, $Zn_2$(TSFTB), is based on the idea that when replacing TTF by Tetraselenafulvalene (TSF, $C_6H_4Se_4$) the larger $p_z$-Orbitals of Se should lead to an increased orbital overlap and this should result in a larger valence band width.

### 1.2. Describing through-space charge transport in pristine MOFs

Before considering the electronic band structure of $Zn_2$(TTFTB) and how it is affected by changes in the structure of the TTF stack it is useful to consider, how *through-space* charge transport in porous materials can be described. As the situation for TTF stacks shown in Figure 1 is strongly reminiscent of (one-dimensional) organic semiconductors, one can benefit here from the insights generated for this materials class over the past decade(s).[31,51–53] Various models for describing charge carrier transport in OSCs have been proposed over time while their suitability typically depends on various parameters (temperature, degree of disorder, molecular arrangement).[31,51–53] These models range from fully coherent band transport as one limiting case to incoherent hopping transport as the other. Additionally, in recent years the dynamic disorder model has attracted considerable attention and derived approaches proved highly successful in describing charge transport in OSCs.[54–58] The core idea of these models is that the charge carrier dynamics is strongly impacted by low-frequency molecular motions which change the intermolecular electronic couplings and in this way determine the charge transport properties of the material. For all these models the intermolecular electronic couplings between neighboring molecules are essential parameters.[31,51–53]. These electronic couplings are, typically, expressed via transfer integrals t. For hopping-based theories the carrier mobility, $\mu$, is proportional to $t^2$, while it is proportional to t for band-transport models, at least within a simple tight-binding picture.[31,53] The transfer integrals can, for example, be extracted from studying suitably arranged dimers[31,34–37,59] or from fitting tight-binding models to electronic band structures.[60] In fact, within a tight-binding picture, the magnitudes of the transfer integrals associated with the frontier orbitals are intimately related to the widths of the frontier bands. This suggests that a first, fundamental understanding of the correlation between charge-carrier mobilities and the relative alignment of the molecular building blocks of a MOF can be gained from calculating electronic band structures, even in cases in which band transport is not the dominant mechanism. Therefore, in the following, we will primarily analyze such band structures calculated by dispersion corrected density functional theory (with the discussion primarily based on band widths and the derived transfer integrals). We acknowledge that in this way the role of the material's phonon properties (like the occurrence of “killer phonon” modes)[61] is not revealed. Nevertheless, the analysis provides crucial insights into structure-to-property relationships for those electronic MOF properties that determine through-space charge transport.

## 2. Methods

For investigating the structural and electronic properties of the MOFs, the periodic model systems, and the molecular dimers we employed dispersion corrected density functional theory, DFT, which in a recent review has been highlighted as a viable method for gaining an in-depth understanding of the electronic structure of MOFs.[62] The simulations were performed with the FHI-aims code.[63] Exchange and correlation were treated by the PBE functional[64,65] and the Tkatchenko-Scheffler[66] scheme was used as an a posteriori van der Waals correction. We employed the default tight basis sets of FHI-aims for periodic and molecular simulations. Further details on the employed basis functions are provided in the SI. For $Zn_2$(TTFTB) the electronic band structure was also calculated with the hybrid functional HSE06[67,68] to ensure that the introduction of Hartree-Fock exchange has a neglibile influence on the valence band width, as the primary quantity of interest for the present study.

During the geometry optimizations all atomic positions were relaxed until the largest remaining force component on any atom was smaller than 0.01 eV/Å. For all MOF systems a 3x3x3 k-point grid was used for sampling reciprocal space during the geometry relaxations. During the electronic structure calculations a 4x4x4 k-point grid was employed. For both grids the total energy is converged to within less than 1 meV. The smaller grid in the more time consuming geometry relaxations was used to speed up the calculations. For the periodic stacks, we used a 1x1x12 k-point grid, which is already converged even for the smallest system (with the largest reciprocal space vector along the stacking direction). The effective masses were calculated from the (inverse) curvature of the band structure at the top of the valence band in (001) direction to describe transport in the TTF stacking direction. Technically, the band curvature was determined by fitting a cosine function to the dispersion relation E(k) including the 10 k-points closest to the valence band maximum with a spacing between neighboring k-points of 0.005 $Å^{-1}$. A cosine function was chosen for the fit to be consistent with the best suited tight-binding band model for the systems studied here (see below).The structures of the MOFs and the molecular systems were visualized using Ovito (version 3.2.1)[69] the molecular orbitals were rendered using Avogadro (version 1.2.0).[70]

# 3. Results & Discussion

## 3.1 Electronic structure of $Zn_2$(TTFTB) and model system

As a first step, we analyze the electronic structure of $Zn_2$(TTFTB), for which the frontier bands are shown in Figure 2a. Figure 2b contains a zoom into the valence band region. In the following, we will be primarily concerned with bands in ΓA direction, as this corresponds to the stacking direction of the TTF molecules. Moreover, the valence and conduction bands are flat in directions perpendicular to ΓA (with band widths around 1 meV in AL and even more narrow bands in ΓK direction). This indicates that there is virtually no electronic coupling between individual TTF stacks within the MOF. As a consequence, charge transport in $Zn_2$(TTFTB) is essentially one-dimensional. This is supported by measurements by Sun et al, who observed that the electrical conductivity in the direction of the stacks is 2-3 orders of magnitude larger than perpendicular to them.[71]

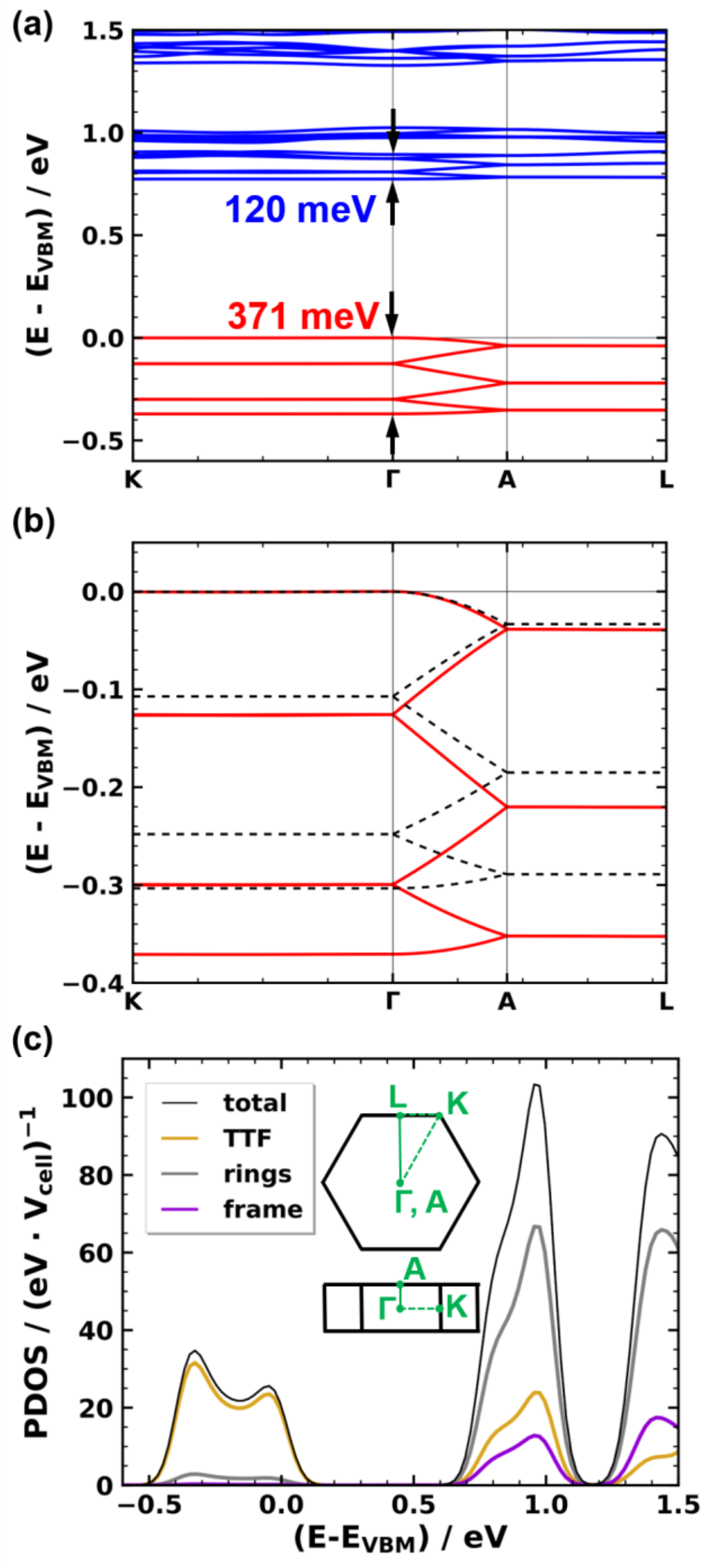

***Figure 2.*** *Electronic structure of the $Zn_2$(TTFTB) MOF and the corresponding TTF model stack: (a) Electronic band structure of $Zn_2$(TTFTB) along selected high-symmetry directions. The energy scale is aligned to the valence band maximum. (b) Zoom into the valence band of $Zn_2$(TTFTB) (solid, red line) and of the TTF model stack (dashed, black line). (c) Density of states of $Zn_2$(TTFTB) projected onto individual parts of the MOF. "TTF" refers to the TTF units, "rings" to the attached phenylene rings and "frame" to the nodes. The total density of states (as the sum of the individual contributions) is given by the black line. A more detailed version of the species projected DOS with separate panels highlighting the contribution of each*

*species is contained in the Supplementary Material. The first Brillouin zone together with the relevant directions in k-space are shown as an inset.*

As far as the ΓA direction is concerned, one can identify a 6 times backfolded band, which is particularly well resolved for the valence band in Figure 2b. This backfolding is related to the crystallographic unit cell (determining shape and size of the first Brillouin zone). It contains 6 TTF-based linkers as translational repeat unit, whose repetition yields the infinitely extended TTF stack. What counts from an electronic point of view is, however, not only the translational symmetry, but also the screw axis in the middle of the TTF stack (see Figure 1). With respect to that screw axis, each TTF molecule has an identical electronic environment. Thus, one can view a single TTF molecule as the "electronic" repeat unit of the TTF stack in $Zn_2$(TTFTB). This is supported by the observation that for the perfectly symmetric structure no band gaps open for the backfolded bands at the Γ and A points. In passing we note that this situation changes, when defects disturb the perfect symmetry, as will be discussed in section 3.4.

As a consequence of a single TTF molecule serving as "electronic" repeat unit, the electronic bands in ΓA direction can be conveniently described by a tight-binding model with a single molecule per unit cell. This notion is confirmed by the data shown in Figure 3, where the dimer derived band widths are compared to those of the actual TTF stacks. Additional validation data are contained in the Supplementary Material. These considerations imply that the width of the valence band between Γ and A is a measure for the electronic coupling between adjacent groups of six TTF units. Conversely, for judging the electronic coupling between neighboring TTF molecules one needs to consider the width of the entire, six times backfolded band, as indicated by the arrows in Figure 2a. Based on the 1D nearst-neighbor tight-binding model mentioned above, the total band width of the 6 times back-folded band then correspons to 4×t (with t representing the intermolecular transfer integral in stacking direction).

On more quantitative grounds, the valence band width, VBW, for the backfolded band amounts to 371 meV in the PBE calculations (400 meV when using the HSE06 functional), as indicated by the arrows in Figure 2a. This is significantly larger than the width of the conduction band, which is 120 meV. This finding suggests that $Zn_2$(TTFTB) is preferentially a hole conductor,[49] which is is in line with the notion of organosulfur compounds like TTF being good electron donors.[72,73]

For a further analysis, we projected the density of states onto individual parts of the MOF (indicated in Figure 1b): the central TTF stack (TTF), the phenylene rings (rings) and the metal nodes (frame). Similar to the findings in Ref [49], this projected density of states (pDOS, displayed in Figure 2c) shows that the valence band is primarily formed from states localized on the TTF stacks. Analyzing the angular momentum resolved species projected DOS in the Supplementary Material shows that these states are of p-character. The phenylene rings contribute only weakly to the valence band. This is consistent with the notion that the valence band describes a *through-space* pathway for holes along the helical stack of TTF molecules. In contrast, for the conduction band, one can see non-negligible contributions from the phenylene rings as well as from the metal nodes. This means that electron transport also involves other parts of the MOF besides the central TTF stack.

These data suggest that especially hole transport (which is particularly relevant for TTF-based systems; see above) should be described well by the model system described in section 3, which consists of an isolated TTF stack. To test this hypothesis, Figure 2b compares the valence band structure of the actual $Zn_2$(TTFTB) MOF (solid line) to that of the model system (dashed line). Qualitatively, the two band structures are the same. The only apparent difference is a somewhat smaller band width of 303 meV in the model system (which amounts to ~ 82% of the band width of the actual MOF). This leads to a comparably small change in the effective mass at the valence band maximum VBM from 2.05 to 2.40 $m_e$ (with $m_e$ being the mass of a free electron).

We attribute this difference to the overlap of the π-orbitals of neighboring phenylenes in the $H_4$TTFTB linkers in the actual MOF, which is not captured by the model system (see systems TTF and TTFTB in Table 1 and further details in the Supplementary Material). These quantitative differences are, however, rather small compared to the effects discussed below, rendering the TTF stack a useful model system.

Its primary advantage is that in the stack, the relative arrangement of the TTF units can be easily modified by changing the rotation between successive molecules, by modifying the conformation of the individual TTF units, by adjusting the position of the screw axis, or by introducing defects breaking the perfect symmetry of the TTF stack. The impact of such variations on the electronic coupling will be discussed in the following sections.

## 3.2 Dependence of band width and transfer integral on the relative rotation of consecutive TTF units

With a reliable model system at hand, we can now turn to studying the impact of changes in the structure of the TTF stacks on the electronic coupling: It has been shown for a variety of molecular dimers that the relative rotation of neighboring molecules has a tremendous effect on intermolecular electronic couplings.[35,37,39–42] As this effect is a consequence of changes in the orbital overlap upon rotation, one can expect similar effects for the TTF stacks considered in this work. Following the construction procedure for TTF stacks described in section 3 it is apparent that the number of stacked molecules in the unit cell determines their relative rotation. Thus, we consider unit cells containing 1, 2, 3, 4, 5, 6, 8, 12 (corresponding to rotations of 0°, 180°, 120°, 90°, 72°, 60°, 45°, and 30°). In passing we note that 0° and 180° rotations do not produce the same structures as one might expect based on the symmetry of the TTF molecules (at least as long as they are not asymmetrically distorted). This is a consequence of the offset between the position of the screw axis and the molecular center, as shown in Figure 1.

The resulting band structures are shown in the Supplementary Material. They reveal that the cofacial arrangement with 1 repeat unit exhibits the largest valence band width of 1337 meV, corresponding to a transfer integral between neighboring molecules of 334 meV (see data points in Figure 3a and values in Table 1). The band width decreases by a factor of nearly three to 447 meV when considering the system with 2 molecules per unit cell (N=2, or a relative rotation between consecutive TTF molecules of 180°). The band width further decreases for 3 TTF molecules (120° rotation) and reaches a minimum of 180 meV (a transfer integral of 45 meV) for the system containing four molecules in the unit cell (see Figure 3a). Upon further increasing the number of repeat units, the band width again increases slightly (to between 235 meV and 337 meV for N=5, 6, and 8). A steep increase is then observed for 12 molecules per unit cell (i.e., for a relative rotation angle of 30°). Here, a valence band width of 650 meV means a doubling compared to the reference system with N=6, which mimicks the situation in the actual $Zn_2$(TTFTB) MOF. Concommittantly, the effective mass of the holes increases from 0.93 $m_e$ for N=1 to 2.48 $m_e$ for the reference system with N=6 and then drops again to 1,84 $m_e$ for N=12. These considerations show that changing the relative twist between consecutive molecules indeed has a profound impact on the electronic coupling in the TTF stack and that the situation in $Zn_2$(TTFTB) is far from ideal for hole transport.

For obtaining values at intermediate rotation angles we use the dimer model described in section 3. The transfer integrals between neighboring TTF molecules as a function of the rotation angle are then extracted via the fragment orbital method[74] . With the transfer integral at hand and employing a one-dimensional tight-binding model with one molecule as "electronic" repeat unit, one can calculate valence band widths as W=4×t. The evolution of the transfer integrals and the resulting band widths is shown in Figure 3a as solid purple line. At rotation angles at which also data from actual stacks are available, one typically observes an excellent agreement between these data and the dimer results. This suggests that dimer calculations can, indeed be used as an efficient tool for predicting general trends and also for explaining the observations, as we will do below.

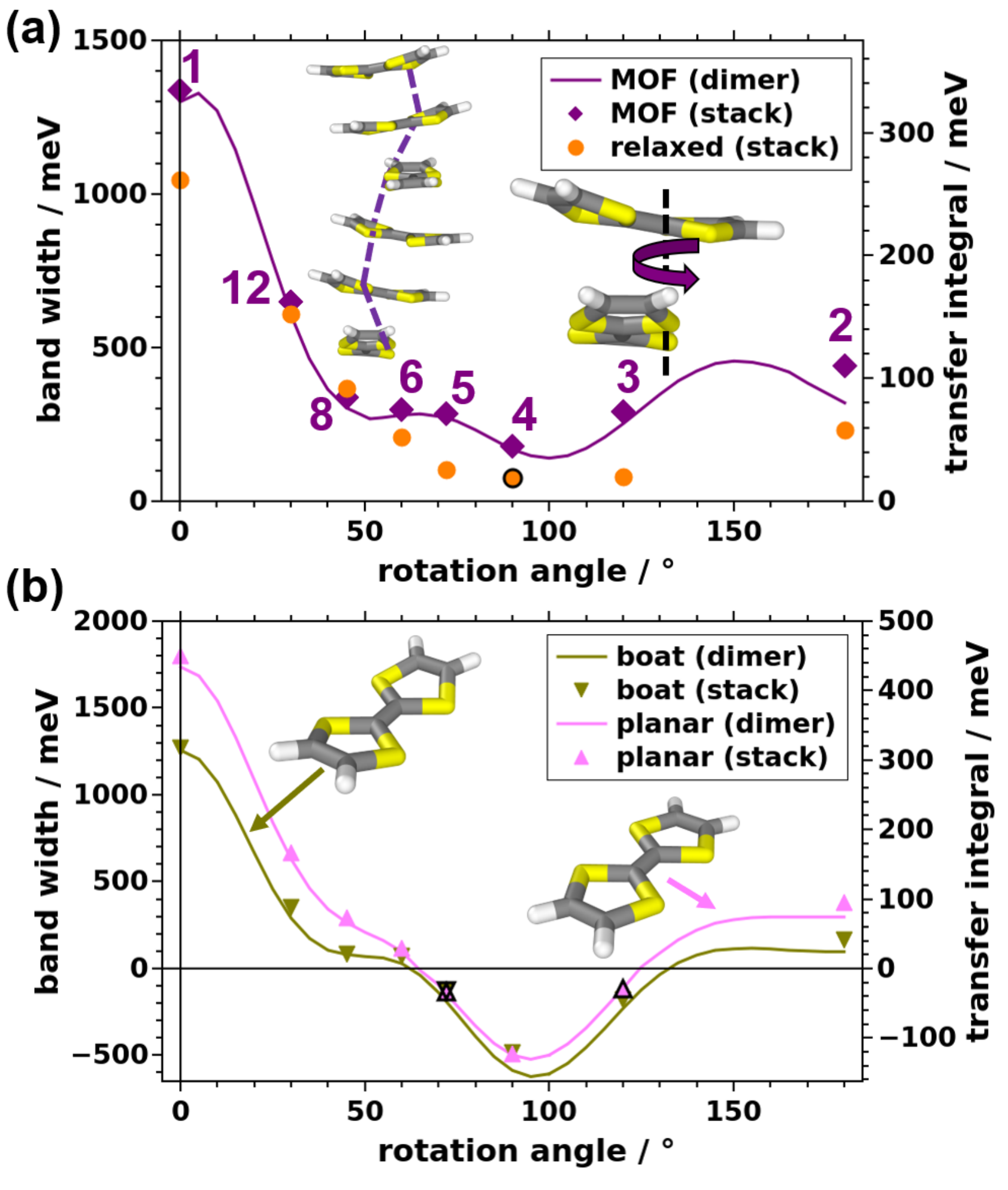

***Figure 3.*** *Evolution of the width of the valence band in (001) direction and of the transfer integrals between neighboring TTF molecules in the stacking direction as a function of the rotation angle between neighboring molecules. Panel (a) shows the situation for stacks and dimers with molecular geometries taken from $Zn_2$(TTFTB) (purple diamonds and line) and for stacks with optimized geometries (orange circles; for details see main text). In Panel (b) the results for fully gas-phase optimized (dark yellow down triangles and line) and for planar molecules (light magenta up triangles and line) are shown. Symbols denote data points*

*calculated for infinitely extended TTF stacks, where the rotation angles are set by varying the number of TTF molecules in each unit cell (see numbers in panel (a)). The solid lines have been calculated for dimers with rotation angles varied in steps of 5° (individual data points not shown). In panel (b) band widths are set to negative values, whenever the signs of the dimer-calculated transfer integrals are also negative. Points marked with a black frame comprise band structures deviating from a simple 1D tight-binding system and are discussed in the Supplementary Material.*

Prior to that, we will, however, discuss the role of the conformation of the molecules within the stack. The data discussed so far have been derived for TTF molecules fixed at the geometries adopted in the $Zn_2$(TTFTB) MOF. We next study the situation, when the geometries of the molecules in each unit are allowed to fully relax, fixing only the positions of the central C=C atoms to maintain the overall structure. As the atoms in the obtained systems can adopt the ideal positions for a given rotation, it also makes sense to compare total energies as a function of the number of TTF molecules per unit cell (see Supplementary Material).

Overall, the evolution of the valence band widths of the relaxed stacks is comparable to that of the MOF-based TTF stacks, although the band widths are consistently smaller in the relexad case (with the exception of the N=8 system; see Figure 3a). In fact, the band widths are particularly small for N between 2 and 5, such that the overall variation between the largest and the smallest band widths amounts to a factor of almost 14. Interestingly, the interaction energy between the TTF molecules in the relaxed stacks are within 35 meV per molecule (i.e., only somewhat larger than $k_BT$) for N between 3 and 12. This occurs despite variations of the band widths (transfer integrals) by a factor of 8. Only for N=1 and 2 the binding energy decreases by 232meV and 121 meV, respectively. This suggests that from a TTF-stacking point of view there is no strong driving force preventing structures with comparably large band-widths (like for N=12). This is in sharp contrast to what has been observed for various acenes and $\alpha$-quinacridone derived molecular crystals.[32,33]

***Table 1.*** *Valence band widths (VBW) and effective masses at the valence band maxima for transport in (001) direction for all considered MOFs (i.e., the parent MOF* $Zn_2$(TTFTB), the MOF with Zn replaced by Cd and the *MOF with S replaced by Se) and for model TTF stacks with 1, 6, and 12 TTF molecules per unit cell (i.e., with rotation angles of 0°, 60° and 30°). The systems TTF and TTFTB listed under MOFs are the stacks extracted from* $Zn_2$*(TTFTB). For the model stacks we compare systems generated with different geometries of the individual molecules. Here, (MOF) refers to TTF conformations extracted from the MOF structure, (relaxed) to geometries relaxed in the stack, (boat) to gas-phase relaxed TTF geometries in boat conformation, and (planar) to planar TTF molecules for which only the x- and y-coordinates have been relaxed in the gas phase.*

| | **MOFs** | |
|---|---|---|
| | **VBW / meV** | **$m^*$ / $m_e$** |
| $Zn_2$(TTFTB) | 371 | 2.05 |
| $Zn_2$(TSFTB) | 641 | 1.05 |
| $Cd_2$(TTFTB) | 333 | 2.21 |
| TTFTB | 373 | 2.10 |
| TTF | 303 | 2.40 |

| | **Model Stacks** | | | | | |
|---|---|---|---|---|---|---|
| | **VBW / meV** | | | **$m^*$ / $m_e$** | | |
| | **N=1** | **N=6** | **N=12** | **N=1** | **N=6** | **N=12** |
| TTF (MOF) | 1337 | 298 | 650 | 0.93 | 2.48 | 1.84 |
| TTF (relaxed) | 1047 | 207 | 609 | 1.89 | 3.02 | 1.86 |
| TTF (boat) | 1269 | 72 | 348 | 1.23 | 7.29 | 4.35 |
| TTF (planar) | 1804 | 117 | 666 | 0.51 | 4.33 | 1.75 |

Notably in the stacks discussed so far (fully optimized and built from molecules in MOF conformation), the TTF molecules are slightly tilted around their long and short molecular axes. To assess the role played by those tilts, we also studied two model systems in which the tilts do not occur. These systems are constructed by starting from a gas phase optimized TTF molecule either in boat conformation (actual minimum structure) or forced to be planar. In the stacks, these molecule are then aligned such that the central C=C bonds as well as all S atoms of each molecule are in a plane perpendicular to the screw axis. The infinitely extended stacks are then constructed following the procedure described in section 1.1. The results for these stacks are complemented by calculations for corresponding molecular dimers. As shown in Figure 3b, the evolutions for both geometries as well as for dimers and infinite stacks are fully consistent. At first glance, they also appear to directly correlate to the data for the MOF-derived structure (purple diamonds and line in Figure 3a). A closer inspection, however, reveals that despite the similar shapes of the respective curves in panels (a) and (b), there is a fundamental difference: The signs of the dimer transfer integrals come out negative for rotation angles Θ between ~65° and ~125°. For the sake of consistency, in that range of rotation angles, we also plot the band widths with a negative sign. This sign cange also indicates that in the bands the wavefunctions at the band maxima and band minima have switched. This zero-crossing of transfer integrals and band widths has a profound impact on charge transport properties. For the latter, in systems like the present one primarily the absolute value of the band width counts. Therefore, for the structures with TTF molecules not twisted around their long and short molecular axes (i.e., for the data shown in Figure 3b), the carrier mobility in stacking direction is expected to be close to a local maximum for the N=4 system (rather than close to the global minimum, as for the cases shown in Figure 3a). Instead, the valence bands become completely flat for rotation angles around ~65° and ~125°. I.e., now for these rotation angles charge transport in the stacking direction will be essentially blocked.

The origin of the zero-crossing (and, in fact, the entire evolution of the transfer integrals with rotation angle) can be explained by the shapes of the involved molecular orbitals. This is most straightforwardly seen by considering the bonding and antibonding linear combinations of the highest occupied molecular orbitals (HOMOs) of the two TTF molecules in the dimer calculations. They can be derived from linear combinations of the HOMOs of indivifual TTF molecules and (for centrosymmetric systems) their splitting determines the magnitude of the transfer integral.[31] The evolution of the orbital shapes and orbital energies with rotation angle is exemplarily shown in Figure 4 for the dimer consisting of planar molecules.

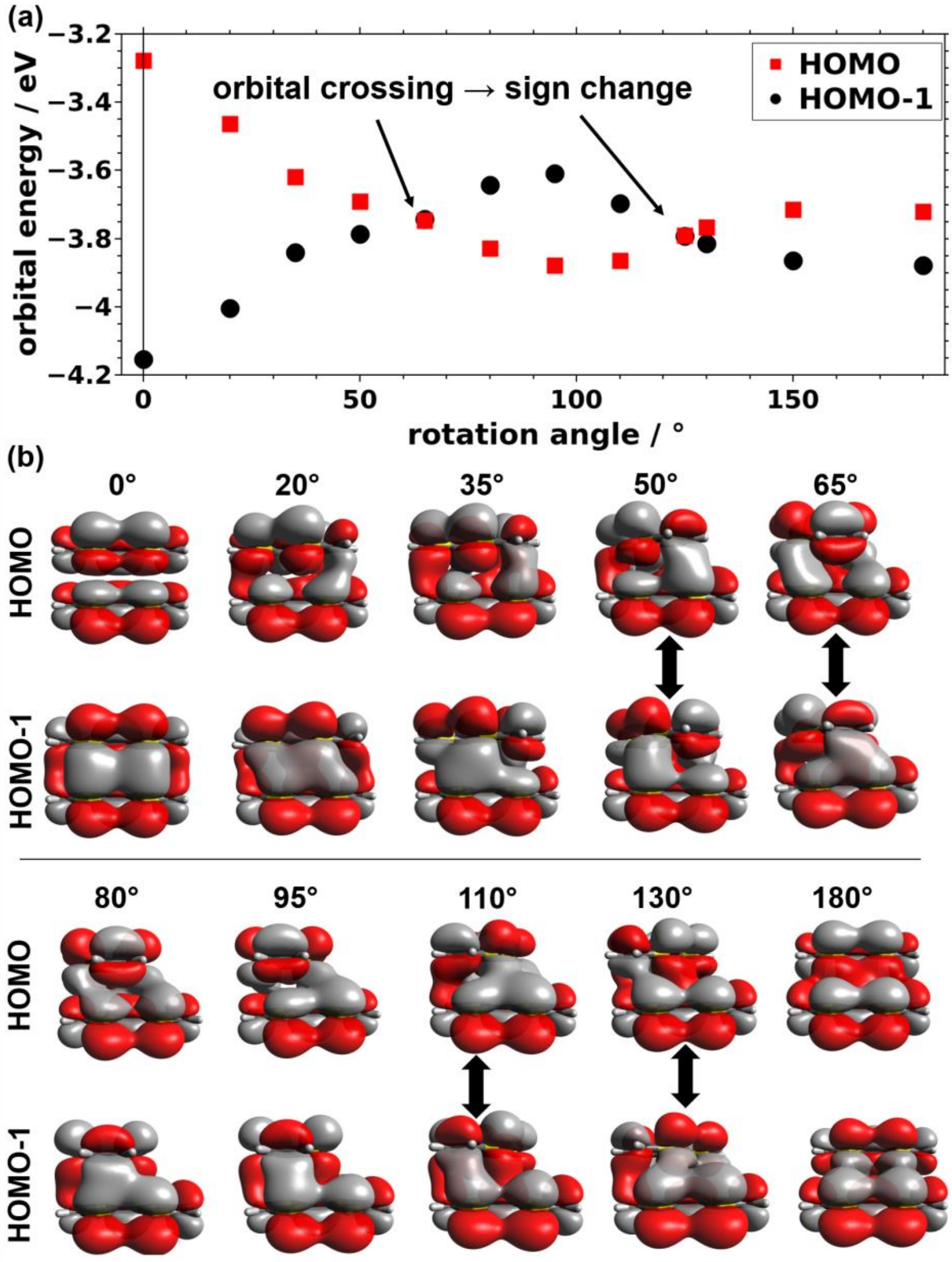

***Figure 4.*** *Orbital energies of the planar TTF dimer as a function of the rotation angle as well as corresponding plots showing the respective molecular orbitals. In cases, in which the two frontier orbitals have change their order, this is highlighted by thick vertical blue arrows in panel (b).*

For the cofacial arrangement of the molecules the antisymmetric linear combination of the TTF HOMOs is lowest in energy and the symmetric linear combination is highest (Figure 4b, 0°). This is exactly what one would expect considering the fully bonding nature of the hybrid orbital in the antisymmetric case (non-zero value of of the wavefunction between molecules or even a local maximum) and its fully antibonding nature in the symmetric case (vanishing wavefunction between the molecules). Upon increasing the rotation angle, the HOMO-1 gets increasingly destabilized and the HOMO gets increasingly stabilized, which reduces their energetic splitting and concommittantly the associated transfer integral. This can be understood by the appearance of antibonding contributions for the antisymmetric and bonding contributions for the symmetric linear combinations. At a rotation angle of 65° both linear combinations display equal amounts of bonding and antibonding regions. Consequently, the two orbitals are essentially isoenergetic, resulting in a vanishing transfer integral. Upon further increasing the rotation angle, the nature of the HOMO and HOMO-1 is switched, which explains the change in sign of the transfer integral. The stabilization of the originally antisymmetric linear combination of the molecular orbitals is maximized at a rotation angle of 95° and the trend reverses for systems with larger rotations. In passing we note that the reason for the much smaller HOMO to HOMO-1splitting at 180° compared to the cofacial situation (i.e. 0°) is the reduced spatial overlap of the molecules following from the screw axis not coinciding with the center of the TTF molecules (see Figure 1).

Similar trends are observed for the other three molecular conformations. The lack of a zero-crossing of the band widths and transfer integrals for the MOF-derived and optimized geometries (i.e., the systems shown in Figure 3a) arises from the fact that due to the twisting of the molecules around the long and short molecular axes certain regions of neighboring molecules get particularly close. This strongly amplifies the contributions of these regions to the orbital energies such that the cancellation effects discussed above do not occur any more.

Finally, it should be noted that, of course, also the location of the screw axis relative to the center of the TTF molecules strongly impacts the wavefunction overlap between neighboring TTF molecules and, consequently, the effects discussed above. To quantify that, we calculated planar model dimers stacked right on top of each other with the screw axis connecting the centers of the planar molecules. The resulting evolutions of the band widths and transfer integrals are shown in the Supplementary Material. The data points for a rotation angle of 0° are the same as for the planar geometry in Figure 3b and one again observes a decrease of the coupling with increasing rotation angle. The absolute magnitude of the coupling is minimized for an intermediate rotation (~50°). In contrast to the situation for the off-center screw axis in Figure 3b, for a centered screw axis one never observes a change in the signs of the band widths and transfer integrals. This can again be directly traced back to the dimer orbitals, which never change their order, as shown in the Supplementary Material.

### 3.3 Interplay between intermolecular distance and band width

The above observations show, how the band width of the TTF stack depends on the relative rotation angle of neighboring molecules around the screw axis of the stack and on the tilt of individual molecules around their long and short molecular axes. Modifying either of those structural parameters has the potential to increase the width of the valence band, yielding improved hole transport properties. Another structural parameter that changes the intermolecular electronic coupling is the distance between neighboring TTF molecules, which for periodic stacks can simply be changed by modifying the stacking distance between the molecules. In periodic calculations this can be set by the extent of the unit cell in the direction of the screw axis. Therefore, we next changed the stacking distance by ±0.1 Å per molecule in the unit cell for the infinitely extended TTF stack and equally modified the distance in the corresponding dimers (shifting one of the molecules by ±0.1 Å in the direction of the screw axis). The results of these modifications are shown in Figure 5a for molecules adopting the same conformation as in the MOF and in Figure 5b for planar molecules. Not unexpectedly, the band width typically increases upon decreasing the inter-molecular distance and vice versa. The data in Figure 5 also show that typically the absolute change of band witdths and transfer integrals with stacking distance is more pronounced for situations in which these quantities are already large to start with. This can be rationalized based on the discussion of Figure 4 in the previous section: In cases in which bonding and antibonding contributions for certain hybrid orbitals largely cancel, this situation is not fundamentally modified upon changing the stacking distance. Conversely, when hybrid orbitals are either fully antibonding or fully bonding (as in the case of the cofacial dimer), changing the stacking distance has a maximal impact.

Therefore, in this case the change of the band width becomes as large as +226 meV (or 17%) for a decrease of the stacking distance by -0.1 Å. Interestingly, for molecules in the MOF conformation changing the stacking distance has virtually no impact for a rotation angle around 90°. We attribute this to a subtle interplay of different relative changes of bonding and antibonding contributions resulting from different local distances due to the twists of the molecules around their long and short axes.

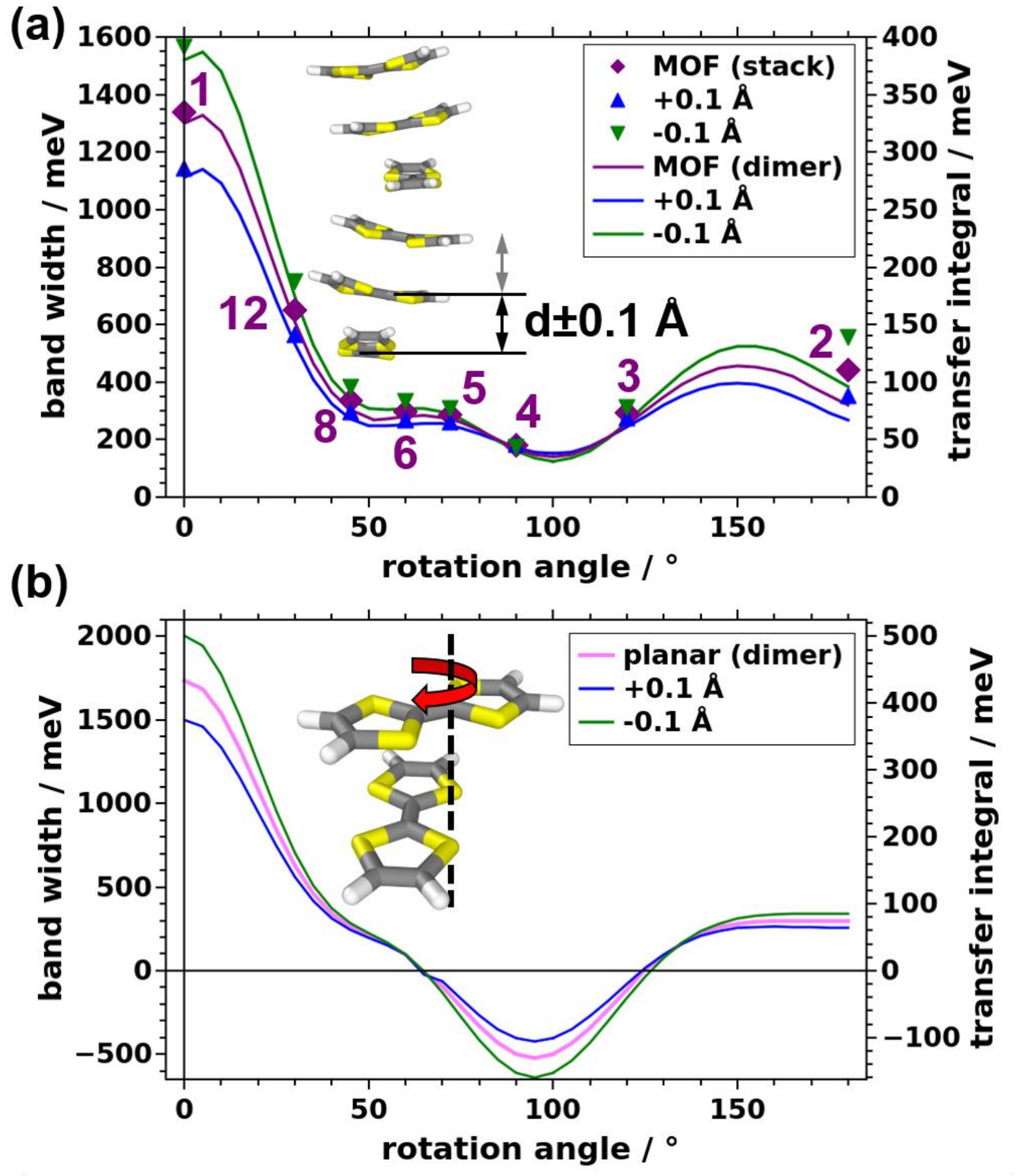

***Figure 5.*** *Evolution of the band width and the transfer integral with the rotation angle as a function of the stacking distance. (a) Results obtained for TTF stacks and corresponding dimers based on the MOF geometry. (b) Dimer results based on the planar TTF conformation.*

Notably, for the N=6 stacks, which directly mimicks the stacking of the TTF molecules in actual $Zn_2$(TTFTB) MOF the increase of the valence band width for a reduction of the stacking distance by 0.1 Å amounts to only 33 meV (~11%). This is insofar relevant, as it has been reported that changing the stacking distance for TTFTB based MOFs results in massive changes in the measured electrical conductivities.[49] Especially, when changing the cation employed in the synthesis from $Zn^{2+}$ to $Cd^{2+}$ an increase of the electrical conductivity by two orders of magnitude was observed. This was originally attributed to the lowered S…S distances for neighboring TTF units, which decreased by 0.119 Å.[49] Such a massiv change in conductivities cannot be explained by the trends discussed above. This raises the question, whether there are relevant structural changes between $Zn_2$(TTFTB) and $Cd_2$(TTFTB) MOFs other than a change in the stacking distance. Therefore, we compared the full electronic band structures of $Zn_2$(TTFTB) and $Cd_2$(TTFTB), but also in this case the changes in band widths and effective masses for the valence band are comparably small, as shown in Table 1. The electronic band structure for $Cd_2$(TTFTB) is shown in the Supplementray Material. In fact, the valence band width is even smaller in $Cd_2$(TTFTB) compared to $Zn_2$(TTFTB).

A different approach for increasing the valence band width could be to increase the orbital overlap by exchanging TTF with Tetraselenafulvalene ($C_6H_4Se_4$). In our calculations this results in an increase of the valence band width by a factor of almost 2 (from 371 meV to 641 meV). Considering that we observe only minor structural changes between the S-based $Zn_2$(TTFTB) and the Se-based $Zn_2$(TSFTB), we attribute that to the larger spatial extent of the $p_z$-orbitals of Se, which result in an increased wave-function overlap. Still, also such chemical modifications do not change hole mobilities by orders of magnitude.

An additional factor that would influence the mobility of holes would be changes in the vibrational properties of the MOF, which impact charge transport through dynamic disorder effects.[53,54,61] Such effects are not explicitly considered here, but it is hard to imagine that they could lead to orders of magnitude changes in transport properties between the materials considered above $Zn_2$(TTFTB) and $Cd_2$(TTFTB).

There could, however, be several other explanations for the above mentioned differences between $Zn_2$(TTFTB) and $Cd_2$(TTFTB). First, one has to keep in mind that in [49] the authors did not report carrier mobilities, but electrical conductivities (which to date is still common for metal-organic frameworks).[71] These are crucially impacted not only by carrier mobilities, but also by the densities of mobile carriers. As far as the latter are concerned, it has been argued in the past that they can be massively impacted by the nature of the metal ions in the nodes.[12,75] Another factor massively changing free carrier concentrations in any type of semiconductor are extrinsic impurities (i.e., dopants).[12,76–80]

Besides chemical imperfections influencing the carrier concentration, also structural imperfections can have a tremendous impact, in this case also on the carrier mobilities. The impact of some of these imperfections on the electronic coupling in $Zn_2$(TTFTB) type systems is discussed in the following section.

## 3.4 Role of defects

A consequence of the flat electronic bands along reciprocal space directions perpendicular to the TTF stacks (see section 3.1) is that charge transport is essentially one-dimensional. It is well established for molecular semiconductors that transport in 1D systems is severely affected by either static or dynamic disorder.[54] This is not surprising, considering that an "obstacle" along a 1D transport path cannot be simply bypassed via neighboring sites. In the context of MOFs, it has, actually, been found that defects can lead to bands with almost no dispersion.[81] For the present systems, we considered several types of static defects. As mentioned above, dynamic disorder caused by vibrations of the MOF lattice is not considered here, although the defects discussed in the following in some sense also mimic what could happen as a function of the thermal motion of the MOF consitituents.

The static defect with the most dramatic consequences is a missing linker defect. We realized such a defect by removing one TTF linker from either the $Zn_2$(TTFTB) unit cell or from the corresponding model stack (see Figure 6a). To describe pair formation, we displaced one molecule in the unit cell such that it moved towards one of its neighbors by -Δd and away from the other neighbor by +Δd (see Figure 6b). A "displaced molecule" defect is characterized by one of the molecules in the unit cell shifted from its equilibrium position along a vector parallel to the xy-plane (Figure 6c). It is well known for OSCs that the changes in the intermolecular interactions induced upon variations of the involved molecule's relative displacements depend on the actual shift direction.[33–36,38] However, in the present contribution we chose displacements along the x-direction (Figure 6b) as representative examples highlighting the potential impact of such defects. For the final defect that is explicitly considered, the “misrotated” molecule case, the rotation angle of one of the molecules in the unit cell is changed by a value of $\Delta\Theta$ (Figure 6d). Considering the structure of the MOF and identifying the degrees of freedom which of each TTF moiety, one could actually identify several more structural defects. Examples are tilts of the molecules around the long and short molecular axes, changes in the bending of the molecules, or torsions around the central C=C bonds to name a few. Therefore, missing linker, pair formation, “displaced molecule” and misrotation primarily serve as instructive examples for the possible impact of such structural defects on the electronic structure of the systems. Notably, the qualitative impact of all of the considered defects on the electronic structure of the model stacks is similar. They cause a loss of the symmetry around the $6_5$ screw axis in the middle of the stack. Consequently, also the notion of a single TTF molecule as the “electronic” repeat unit of the stack no longer applies. In the band structure, this results in an opening of gaps at the Brillouin zone boundary and at the $\Gamma$ point (see Figure 7 for the missing linker, displaced and misrotated molecule defects). Thus, for the defective structures it is not sensible to report the width of the 6 times backfolded valence band and we will instead focus on the effective masses at the valence band maximum. Additionally, in the spirit of hopping transport, we will report the smallest transfer integrals between neighboring molecules found in all inequivalent dimers extracted from each of the defective TTF stacks.

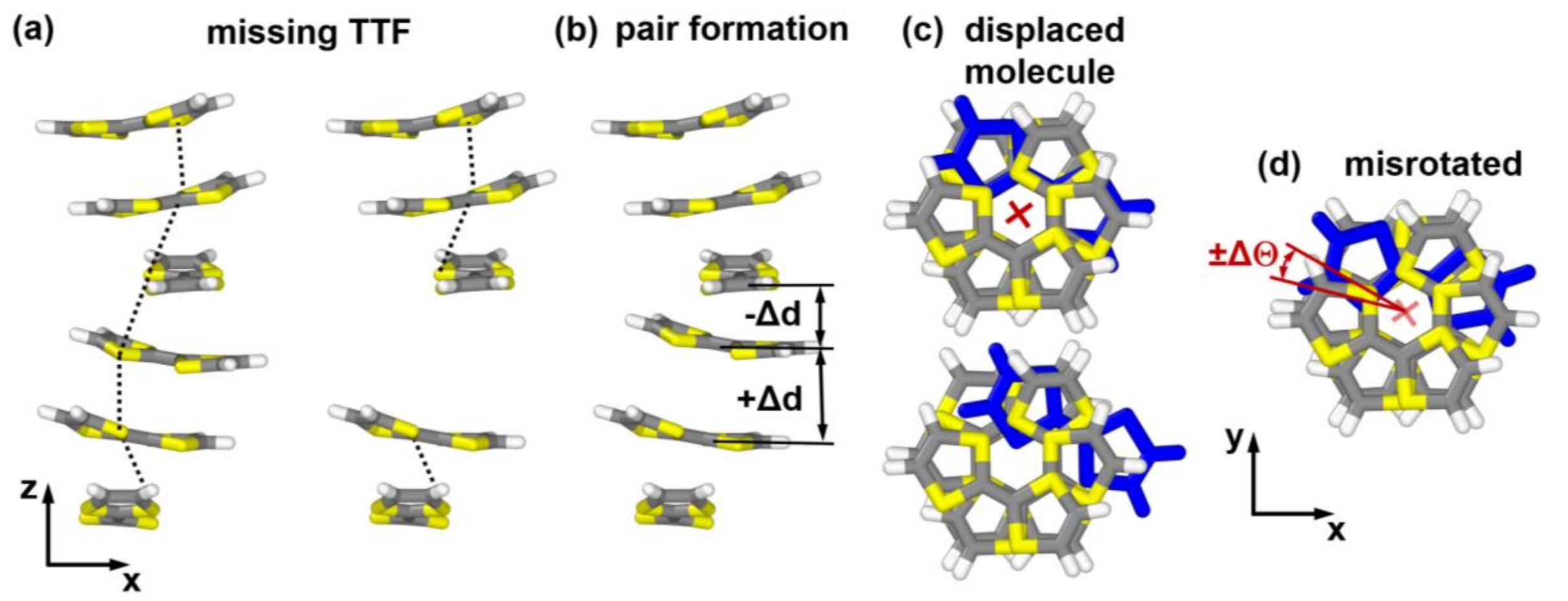

***Figure 6.*** *Instructive examples for possible stacking faults in TTFTB based MOFs. (a): ideal model stack plus system with a missing TTF molecule; (b) Structure of the system upon pair formation between neighboring TTFs; (c) Displaced molecule defect realized by displacing one molecule along x; (d) Structure of a misrotated molecule defect. The coordinate system for the systems in panels (c) & (d) is shown as an inset. It should be noted that for inifinitely extended stacks due to the employed periodic boundary conditions a defect occurs in every unit cell.*

The missing linker defect has the most dramatic impact. It results in essentially flat bands (see Figure 7a), the minimum transfer integral drops to 1 meV and the effective mass skyrockets to 22 $m_e$. This shows that such a defect nearly stops charge transport along the affected TTF stack.

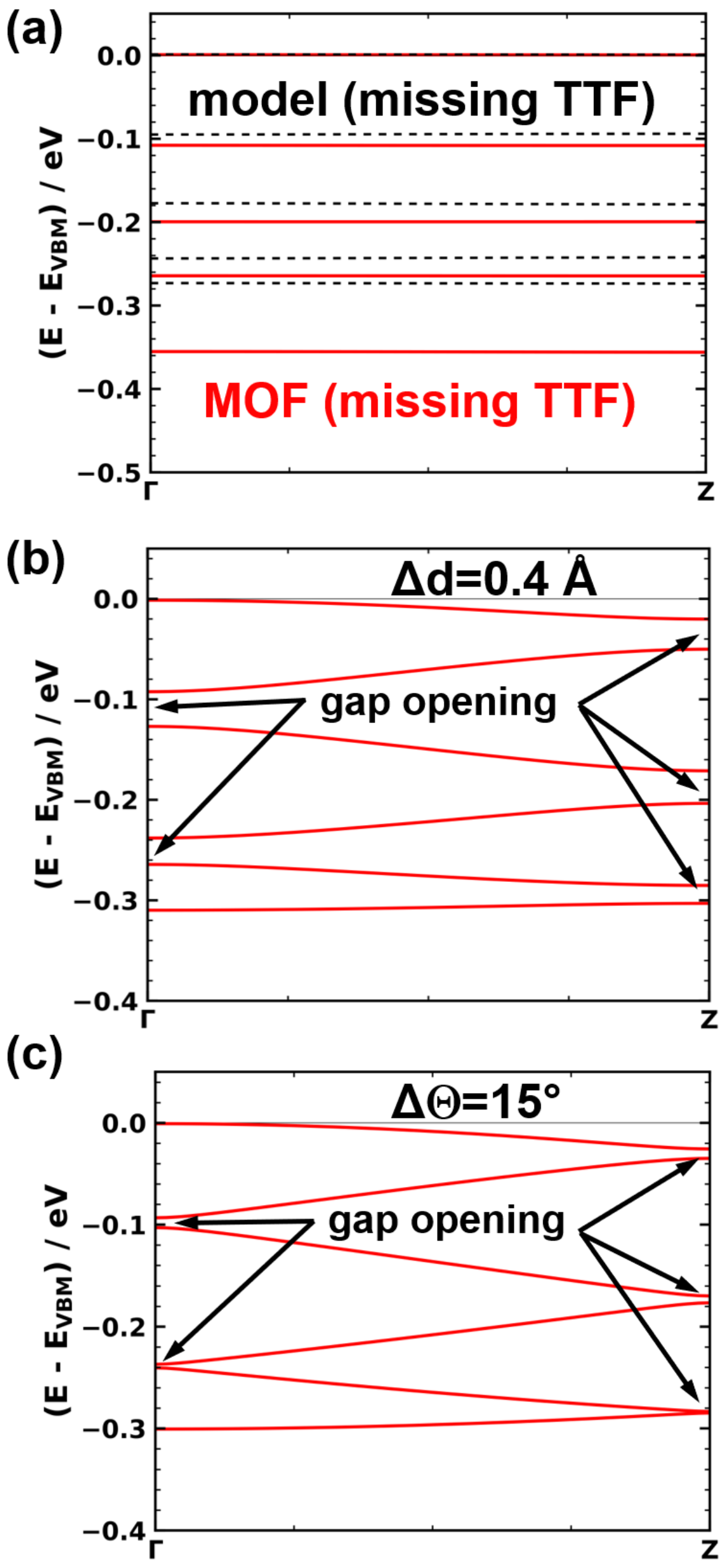

***Figure 7.*** *Electronic structure of the defective systems. (a) Electronic band structure in stacking direction for Zn*$_2$*(TTFTB) with a missing linker defect (red solid line). The results for the corresponding TTF stack are shown as dashed black line. (b) Electronic band structure for the model TTF stack with a pair formation defect with a dimerization of* $\Delta d$*=0.4 Å. (c) Electronic band structure for the model TTF stack with a misrotation defect of* $\Delta\Theta$*=15°.*

Also pair formation and displaced molecule defects result in an increased effective mass and a decreased transfer integral (with the exception of a minor decrease of m* for a very small dimerization of $\Delta d$=0.05 Å, which is in the range of the uncertainty of the fitting procedure). The magnitude of the change increases with increasing displacement, as shown in Figure 8. Additional data on the impact of the other defects can be found in the Supplementary Material. As a consequence, charge transport is hindered within defective TTF stacks. Interestingly, for pair formation as well as for the displaced molecule defect the minimum transfer integral decreases almost linearly with the displacement, while the effective mass experiences a roughly quadratic increase. The latter is more pronounced in the displaced molecule case. The overall impact of these defects is, however, rather moderate (especially compared to the missing-linker case). For example, a lateral displacement of a TTF molecule by a rather sizable distance of $\Delta x$=0.3 Å leads to an increase of $m^*$ by a moderate 0.42 $m_e$ (or 17%).

In this context it, however, has to be considered that there are many other degrees of freedom that can result in molecular displacements, which change the effective masses and transfer integrals. Additionally, several defects might occur simultaneously, further worsening the situation. Nevertheless, the above considerations suggest that for changing the carrier mobilities by orders of magnitude, mere displacements of molecules will typically not be sufficient. Such massive changes in the transport properties will, thus, typically require more serious defects like missing linkers, as discussed earlier in this section.

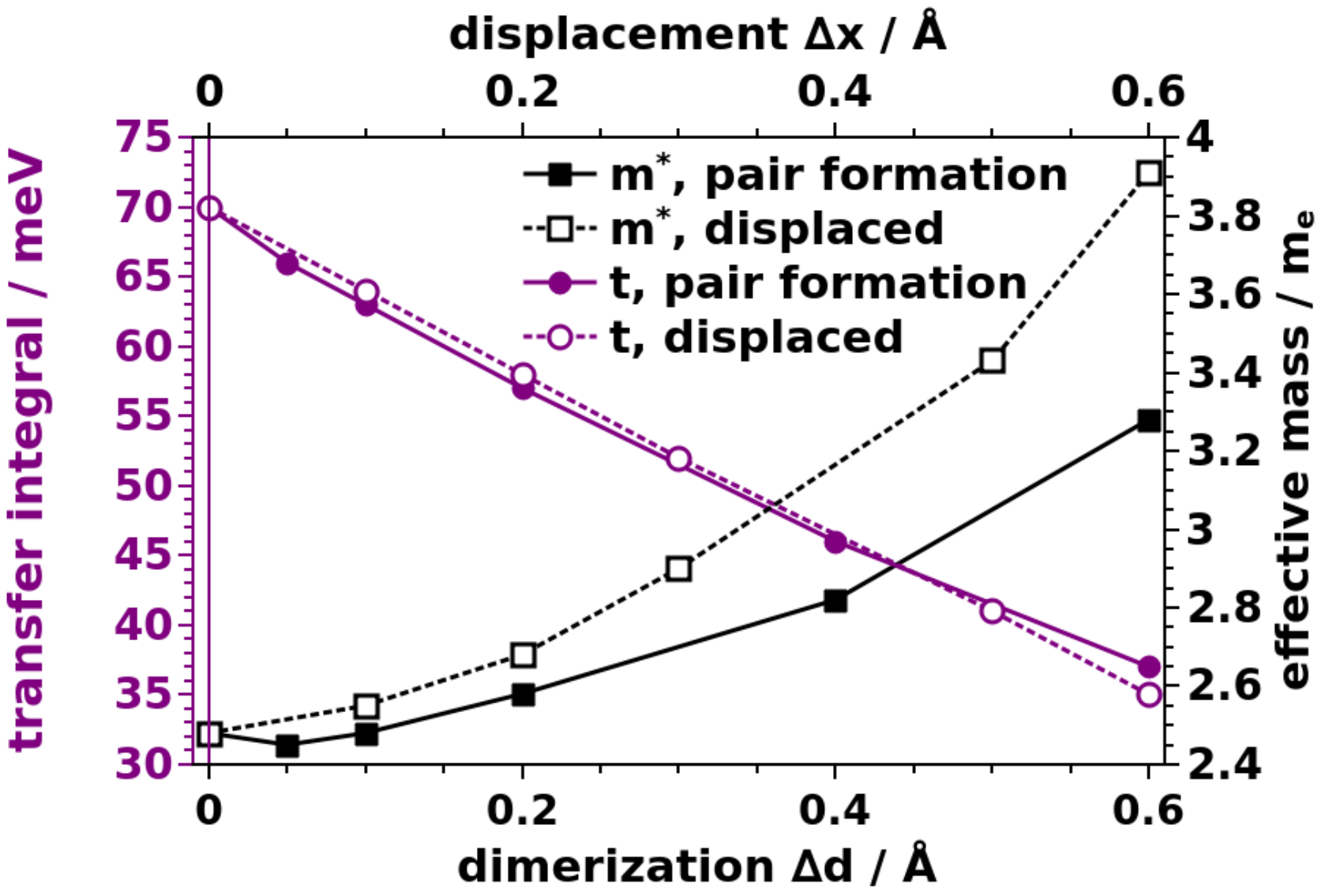

***Figure 8.*** *Evolution of the effective mass and the smallest transfer integral of the model TTF stack as a function of the dimerization $\Delta d$ and displacement $\Delta x$.*

# 4. Conclusions

The present paper describes a variety of aspects concerning through-space charge transport in metal-organic frameworks in general and tetrathiafulvalene-based MOFs in particular. First, it is shown that the electronic band structure of the helical TTF stack contained in $Zn_2$(TTFTB) largely determines the valence band structure of the entire MOF. In fact, we find that the electronic bands perpendicular to the TTF stacking direction are essentially flat. This highlights the negligible electronic coupling between neighboring stacks and establishes that $Zn_2$(TTFTB) is a truly one-dimensional conductor. In the perfectly periodic MOF with six molecules in the crystallographic unit cell the valence band is backfolded six times (without any gaps at the Γ-point or at the Brillouin zone boundary). This suggests that the symmetry element relevant for the electronic structure of the MOF is the 6-fold screw axis parallel to the stacking direction. Therefore, a single TTF molecule acts as "electronic" repeat unit of the MOF with the consequence that the electronic parameter determining charge transport in $Zn_2$(TTFTB) is the transfer integral between two neighboring TTF molecules.

This permits the use of stacks with varying numbers of molecules in the crystallographic unit cell to study the impact of the relative rotation of the TTF molecules. It turns out that decreasing the rotation angle of neighboring TTF molecules compared to the parent $Zn_2$(TTFTB) system significantly increases the valence band width, while increasing the rotation in a 4-TTF per unit cell stack yields a significantly reduced electronic coupling. The results are corroborated by simulations on TTF dimers, which also allow us to trace the observations back to the shapes of the hybrid orbitals determining the valence band. Additionally, we find that the actual value of the transfer integral is extremely sensitive to the specific conformation of the TTF molecules. For example, for stacks of flat TTF molecules the electronic coupling essentially disappears for the 60° rotation angles found in $Zn_2$(TTFTB) and the associated transfer integral even changes sign at larger angles.

Interestingly, changes in the relative rotation and molecular conformation of the TTF molecules have a more pronounced impact on the observed band width than "moderate" modifications in the stacking distance, which have been realized experimentally by replacing Zn by Cd atoms in the metal nodes of the MOFs. Thus, we hypothesize that the 2-orders of magnitude increase in the electrical conductivity of $Cd_2$(TTFTB) compared to $Zn_2$(TTFTB) [49] must either be the consequence of significantly modified concentrations of mobile carriers or must be due to changes different defect densities in the two systems.

As far as static defects are concerned, we have, thus, investigated several scenarios including diplaced molecules, molecular pairing along the stack, or misrotations of specific molecules. The impact of these defects turns out to be rather moderate. This, however, changes when considering also missing linker defects, where we find that due to the 1D nature of the TTF stacks such a missing linker is a massive obstacle for charge transport. This, is manifested, e.g., in an increase of the effective mass by a factor of ~10 compared to the perfectly ordered parent MOF.

Overall, these results show that, on the one hand, there is still considerable room for improvement for through-space charge transport in MOFs through clever structural design. On the other hand, the 1D nature of systems like the ones discussed here makes their expected charge-transport properties particularly sensitive to structural imperfections and, thus, extremely dependent on sample quality.

AUTHOR INFORMATION

ORCID

Christian Winkler 0000-0002-7463-6840

Egbert Zojer 0000-0002-6502-1721

**Corresponding Author**

egbert.zojer@tugraz.at

**Author Contributions**

The manuscript was written through contributions of all authors. All authors have given approval to the final version of the manuscript.

**Funding Sources**

TU Graz Lead Project "Porous Materials at Work" (LP-03).

ACKNOWLEDGMENT

The work has been financially supported by the TU Graz Lead Project "Porous Materials at Work" (LP-03). The computational results have been in part achieved using the Vienna Scientific Cluster (VSC3).

SUPPLEMENTARY MATERIAL for

# Strategy for controlling through-space charge transport in metal-organic frameworks via structural modifications

*Christian Winkler*[1] *and Egbert Zojer*[1,]*

[1] Institute of Solid State Physics, NAWI Graz, Graz University of Technology, Petersgasse 16, 8010 Graz, Austria

# 1. Additional methodological details

## 1.1 Overview of basis functions used in FHI-AIMS

***Table S1.*** *Basis functions that have been used for all calculations performed with FHI-AIMS. The abbreviations read as follows: H(nl,z), where H describes the type of the basis function where H stands for hydrogen-like type function, n is the main quantum number, l denotes the angular momentum quantum number, and z denotes an effective nuclear charge which scales the radial function in the defining Coulomb potential.*[63]

| | H | C | S | O | Zn | Se | Cd |
|---|---|---|---|---|---|---|---|
| Minimal | 1s | [He]+2s2p | [Ne]+3s3p | [He]+2s2p | [Ar]+4s3p3d | [Ar]+4s3d4p | [Kr]+4d5s |
| Tier 1 | H(2s,2.1)<br>H(2p,3.5) | H(2p,1.7)<br>H(3d,6)<br>H(2s,4.9) | ionic(3d,auto)<br>H(2p,1.8)<br>H(4f,7)<br>ionic(3s,auto) | H(2p,1.8)<br>H(3d,7.6)<br>H(3s,6.4) | H(2p,1.7)<br>H(3s,2.9)<br>H(4p,5.4)<br>H(4f,7.8)<br>H(3d,4.5) | H(3d,4.3)<br>H(2p,1.6)<br>H(4f,7.2)<br>ionic(4s, auto) | H(2p,1.6)<br>H(4f,7)<br>H(3s,2.8)<br>H(3p,5.2)<br>H(5g,10)<br>H(3d,3.8) |
| Tier 2 | H(1s,0.85)<br>H(2p,3.7)<br>H(2s,1.2)<br>H(3d,7) | H(3p,5.2)<br>H(3s,4.3)<br>H(3d,6.2) | H(4d,6.2)<br>H(4p,4.9)<br>H(1s,0.8) | H(3p,6.2)<br>H(3d,5.6)<br>H(1s,0.75) | | H(4p,4.5)<br>H(4d6.2)<br>H(1s,0.5) | |

## 1.2 Comparison of a simple tight-binding model to the electronic band structure of $Zn_2$(TTFTB)

Here, we investigate whether a simple, one-dimensional tight-binding model with one "electronic" repeat unit can reproduce the 6-times backfolded band structure of the model stacks (shape and band width). This is relevant, as the transfer integrals reported in the main manuscript have been extracted from calculated actual band widths employing that approach. The model reads $E(k) = 2t * cos(k * R)$, with R being the shift between adjacent TTF molecules in the direction of the screw axis, i.e., the distance between electronic repeat units in stacking direction (R = 3.473 Å). From the electronic band structure of the 6 repeat unit TTF stack we extracted the transfer integral as (1/4)×W, with W being the calculated band width. Using W=298 meV we then calculated the electronic band structure of the model. In order to compare it to the 6 times backfolded band structure of the TTF stack we folded the resulting band back into the crystallographic unit cell of the TTF stack containing 6 molecules. Comparing the two band structures (green for model and red for TTF stack in Figure S1a), we find an excellent qualitative agreement. I.e., the simple tight-binding model with a single TTF molecule as "electronic repeat units" provides a physically meaningful model for the valence band of the TTF stack. Also the quantitative agreement is reasonably good, which supports the use of t = W/4 for depicting the trends for the transfer integrals in the main manuscript.

In the actual stack, there might, of course, also be couplings to next nearest neighbors which might play a role. Such couplings could be hyperexchange-like, as we observed it in quinacridone, a prototypical H-bonded organic semiconductor.[60] Therefore, we also considered a slightly modified model, where we considered next-nearest neighbor couplings as well. It reads: $E(k) = 2t * cos(k * R) + b * 2t * cos(2 * k * R)$. The coupling between these sites was estimated to be around 10% of the nearest neighbor couplings (~8 meV) where b actually is a fit parameter (smallest RMSE for b=0.1). A comparison between the bands resulting from this model and the actual TTF stack is shown in Figure S1b. There, we observe a further improved agreement between the simple tight-binding model and the data of the actual TTF stack, especially at the band extrema.

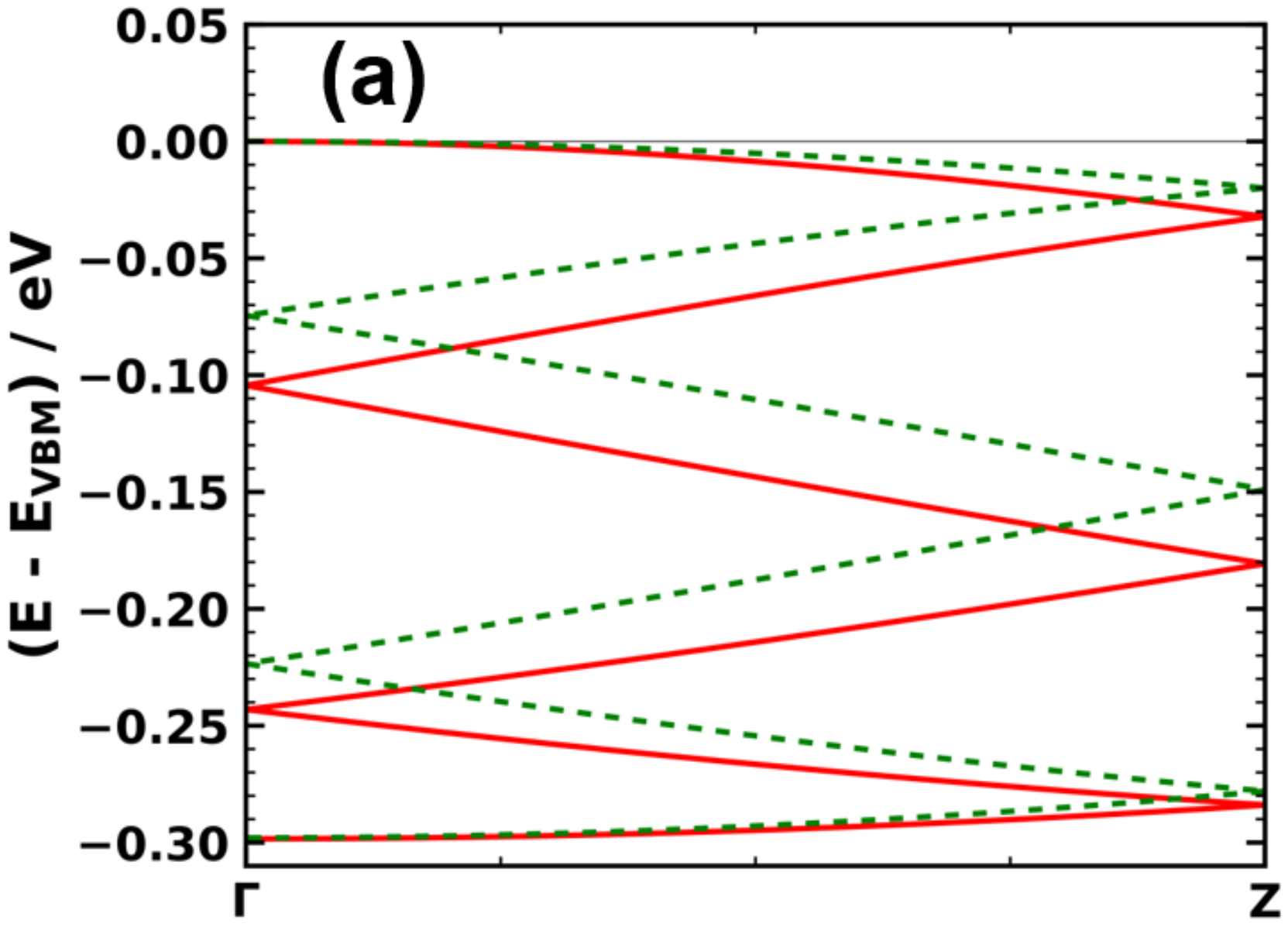

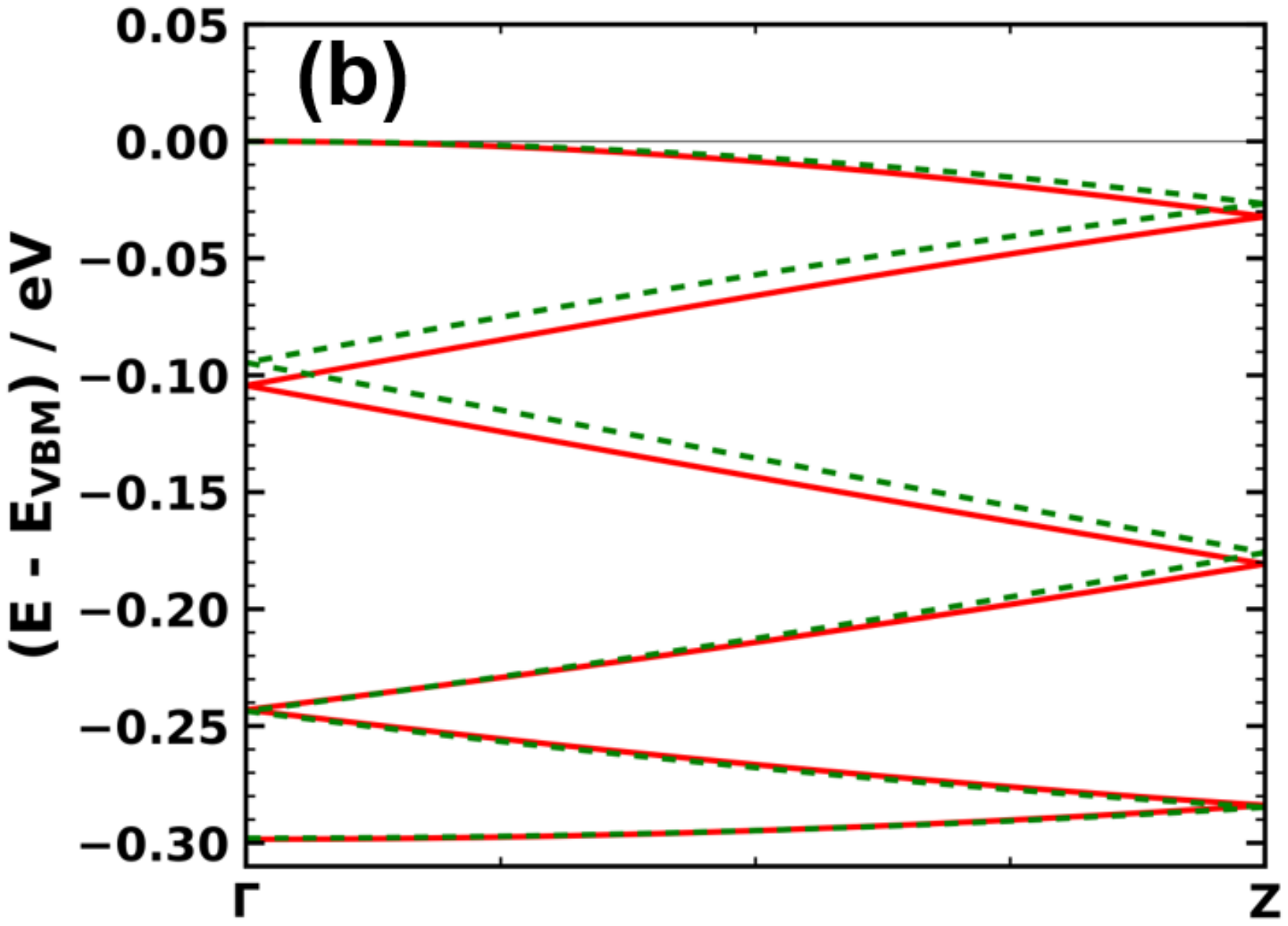

***Figure S1.*** *Comparison of band structures resulting from simple tight-binding models (green dashed line) and the 6 repeat unit TTF stack (solid red line). (a) One-dimensional tight-binding model with only nearest neighbor couplings considered. (b) One-dimensional tight-binding model with next-nearest neighbor couplings also included.*

## 2. Additional data for $Zn_2$(TTFTB) and the TTF model systems

### 2.1 Species projected and angular momentum resolved density of states of $Zn_2$(TTFTB)

In addition to the density of states projected onto sub-systems of the MOF (as presented in the main text) we also considered the DOS projected onto individual species and the DOS resolved by angular momentum (Figure S2). Importantly we find that C and S p-states essentially make up the valence band, while for the conduction band also O p-states and a small contribution from Zn p-states are important.

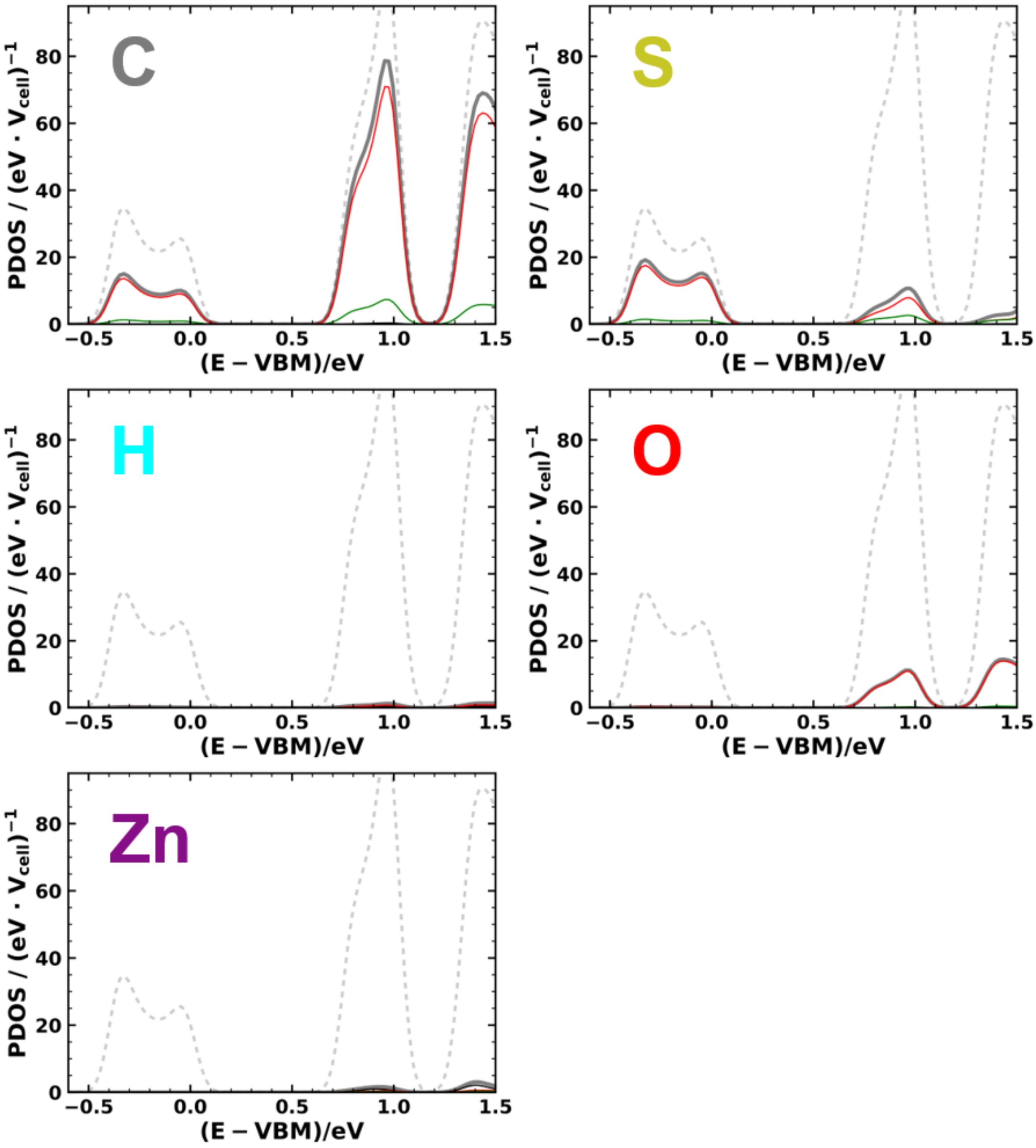

***Figure S2.*** *Species-projected and angular momentum resolved density of states for $Zn_2$(TTFTB). Grey is the total contribution of the individual species, black indicates s-states, red p-states and green d-states.*

### 2.2 Electronic structure of $Zn_2$(TTFTB) calculated with HSE06

To see how the electronic structure of the MOF would be affected by the choice of the actual functional (in particular the treatment of exchange and correlation), we also employed the range separated hybrid functional HSE06 to calculate the electronic structure of $Zn_2$(TTFTB) (Figure S3). As expected for hybrid functionals, this results in an increased band gap (0.773 eV for PBE and 1.483 eV for HSE), which is, however, of no relevance for the discussion in the present manuscript. More relevant is the slightly larger width of the valence band obtained with HSE, but the overall effect is rather minor. These two differences aside, there is virtually no difference between the PBE and the HSE calculations.

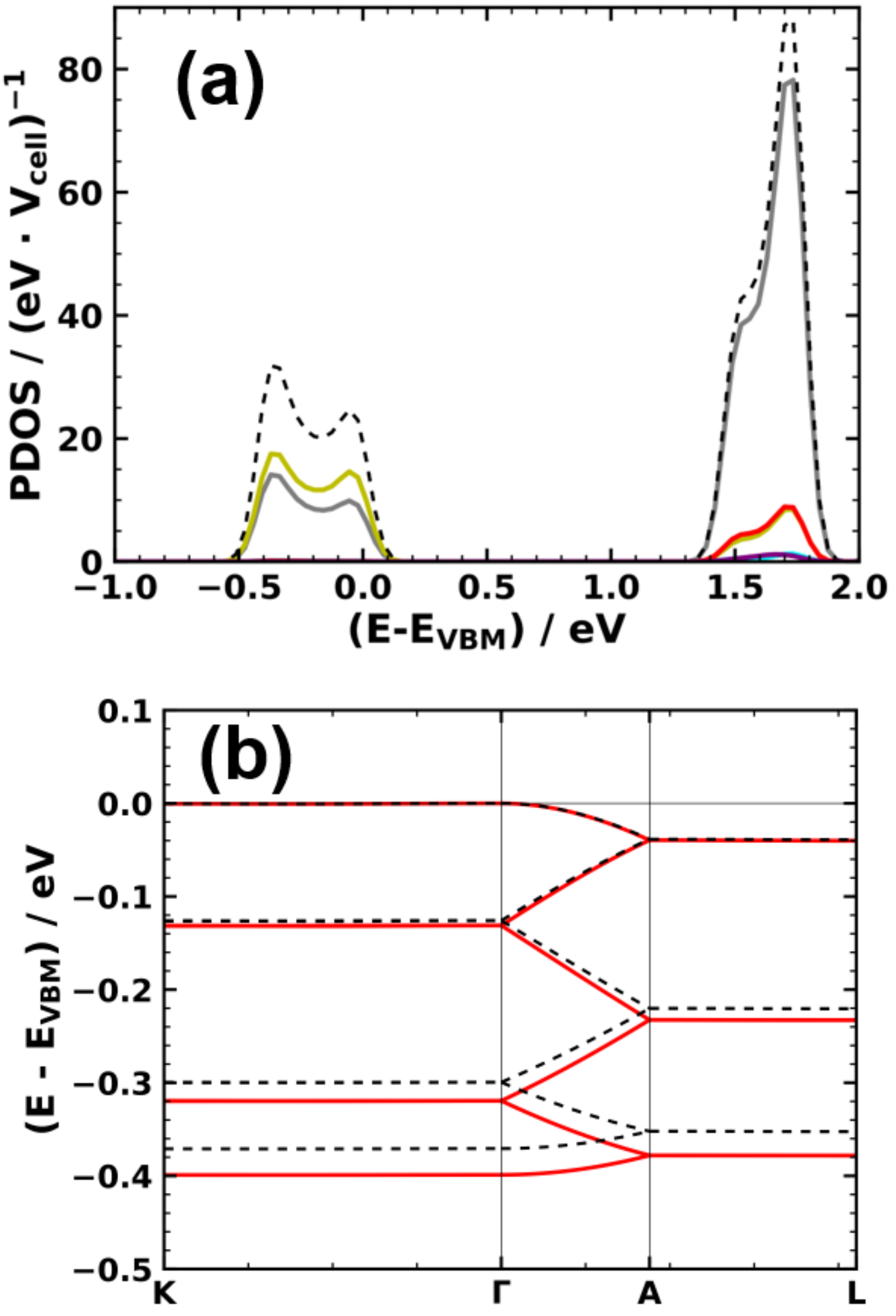

***Figure S3.*** *Electronic structure of $Zn_2$(TTFTB) calculated employing the HSE06 functional and using a PBE-optimized geometry. (a) Species projected density of states (C… grey, H …*

*cyan, O ... red, Se ... yellow, Zn ... purple) and (b) Electronic band structure of the valence band as obtained with HSE06 (red) compared to the PBE result (black, dashed).*

## 2.3 Valence band of $Zn_2$(TTFTB) along additional high-symmetry k-space directions

To investigate whether the valence band is flat for all directions apart from those associated with the stacking direction of the TTF cores (ΓA and LM), we calculated the electronic band structure along a path considering all relevant high-symmetry k-space directions (see Figure S4). Indeed we find that only for directions parallel to the stacking direction of TTF the valence band shows a significant dispersion.

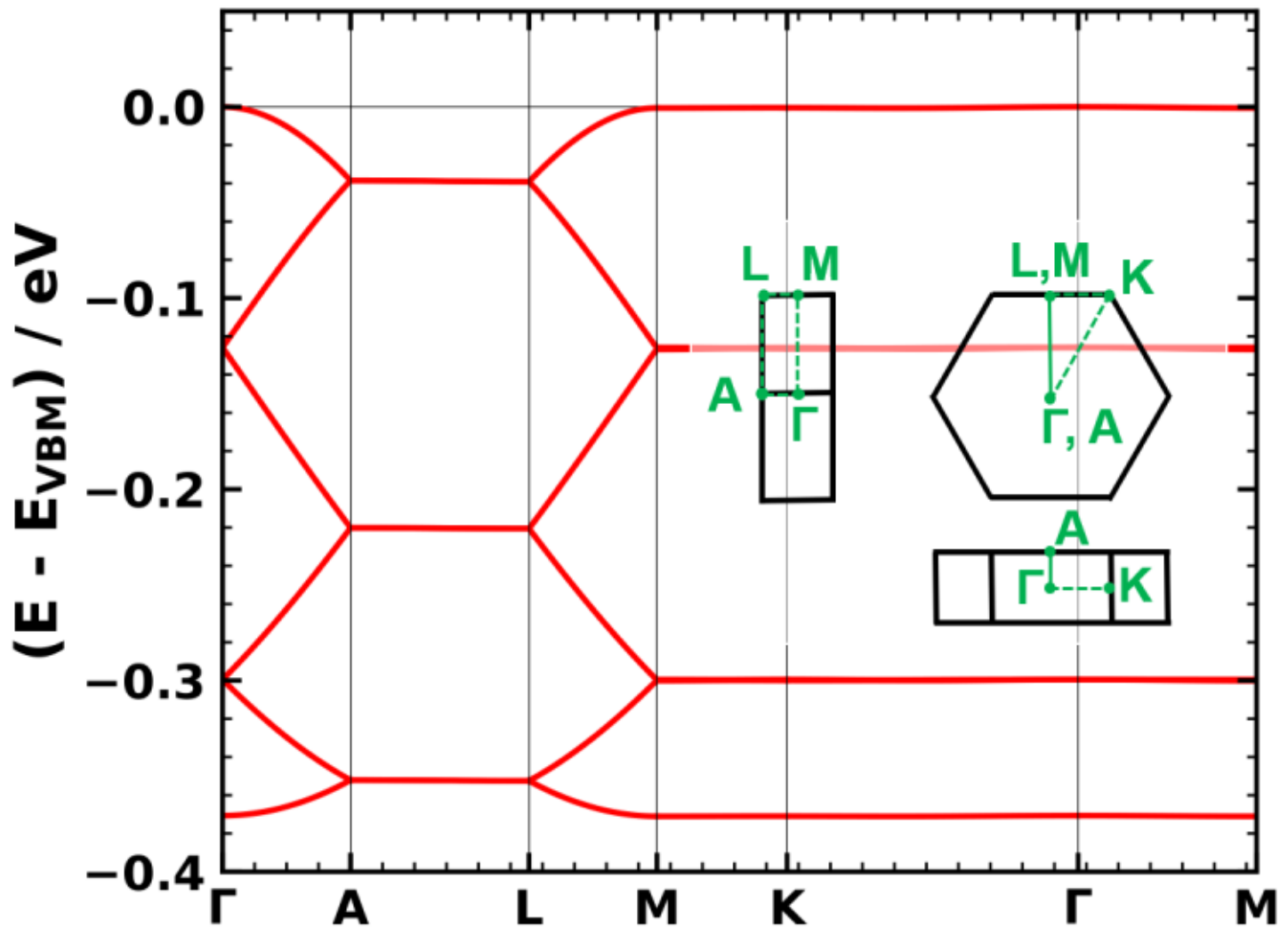

***Figure S4.*** *Electronic band structure for $Zn_2$(TTFTB) for a path covering all relevant high-symmetry directions in the first Brillouin zone. The first Brillouin zone is shown as an inset.*

## 2.4 Geometrical and electronic structure of $Zn_2$(TTFTB) containing water

To investigate the influence of water molecules coordinated to the Zn atoms in the $Zn_2$(TTFTB) MOF, we considered the initial structure extracted from the cif-file from reference [49], removed the solvent molecules and calculated the electronic structure of this system. The used geometry and the obtained electronic structure are shown in the left panel of Figure S5. In the top panel, the structure with the water molecules extracted from the cif file is compared to the structure that has been obtained by relaxing the atomic positions of $Zn_2$(TTFTB) without water (green). One can see that there are only slight differences in the geometric structure of the systems. Considering the electronic band structure of the valence band for this geometry we find a valence band width of 353 meV and a corresponding effective mass of 1.97 $m_e$. Both compare well to the 371 meV and 2.05 $m_e$ obtained for the optimized $Zn_2$(TTFTB) structure without water.

The results for the fully relaxed atomic positions calculated including the water molecules are compared to the relaxed geometry of the system without water (right panel in Figure S5 and Figure S6 a). We find that the central TTF core is hardly affected by the presence of water molecules at the Zn coordination sites, while the phenyl rings indeed show structural changes. As outlined in the main manuscript also these phenyl rings contribute to the valence band. Thus, considering the electronic structure of the relaxed MOF with water we find that this system exhibits a larger band width of 411 meV (1.76 $m_e$) compared to the system without water (see Figure S6 a for band structures). Nevertheless, these changes are quite minor. Therefore, the system without water can serve as a prototypical example for the influence of structural effects on the charge transport properties. Additionally we tested, whether one can use a TTF model system (constructed according to section 3.2 in the main paper) for describing the valence band of the relaxed system with water (dashed black line in Figure S5). Indeed we find that such a TTF stack can serve as a viable model system, only slightly underestimating the resulting valence band width of the actual MOF. For the sake of completeness we also report the species projected DOS of the relaxed $Zn_2$(TTFTB) MOF with water molecules coordinating to Zn in Figure S6.

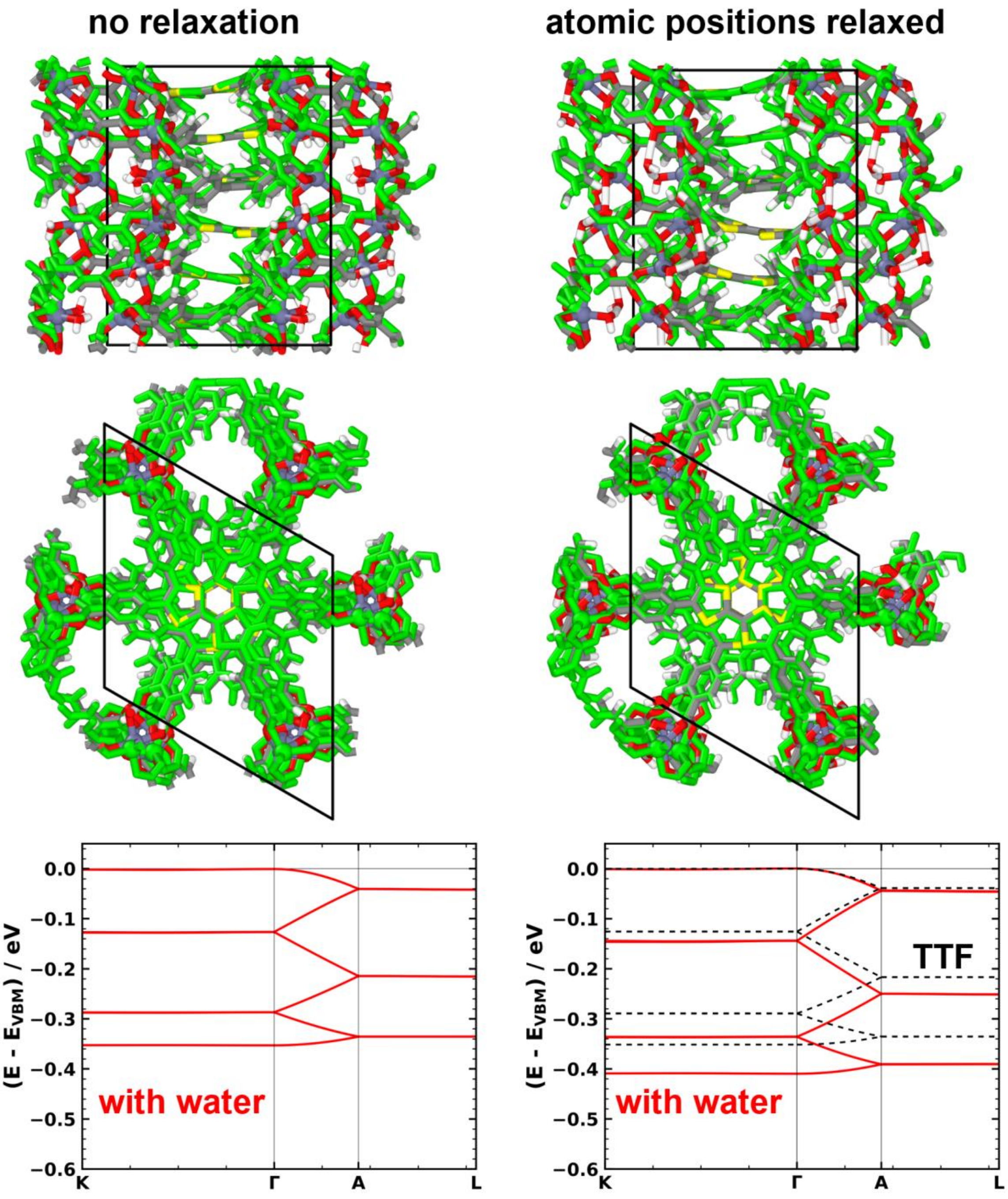

***Figure S5.*** *Geometric structure and electronic band structure of* $Zn_2$*(TTFTB) with and without water molecules coordinated to the Zn metal atoms. Two systems are considered and compared to the structure obtained by relaxing the atomic positions without water molecules present (green): The left panels show data for a structure with water, as reported in literature cif file (i.e., without a further geometry relaxation). The electronic band structure is shown below the crystallographic structure. The right panel shows the structure with water after performing a geometry relaxation of the atomic positions. In the bottom panel the electronic band structure of that system is shown in red. The dashed black line corresponds to the data for the saturated*

*TTF stack including water molecules. The unit cell is shown by the solid black lines in the geometric structures.*

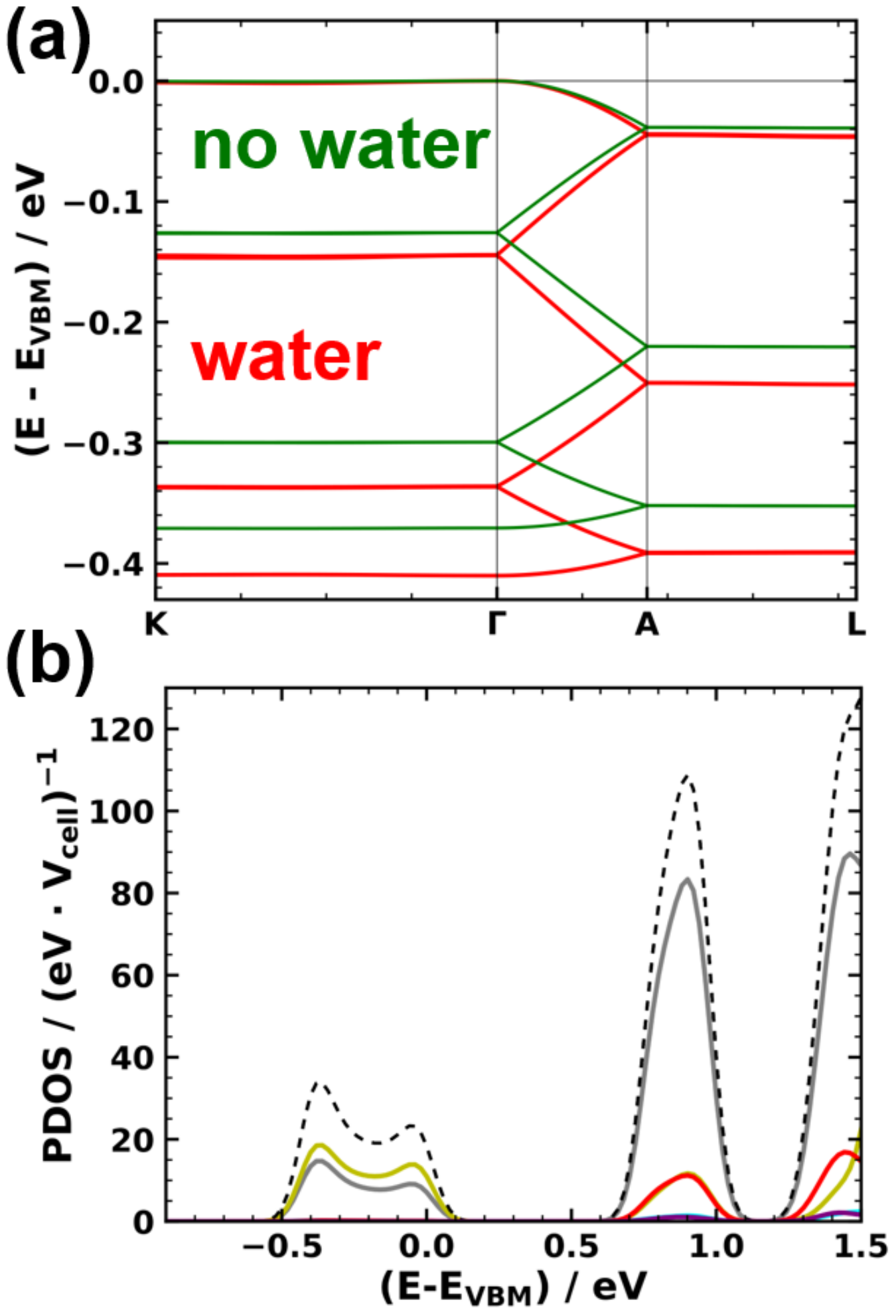

***Figure S6.*** *Electronic structure of Zn2(TTFTB) of relaxed $Zn_2$(TTFTB) with water molecules coordinated to Zn. (a) Electronic band structure of $Zn_2$(TTFTB) with water (red) compared to $Zn_2$(TTFTB) without water (green). (b) Species projected density of states of $Zn_2$(TTFTB) (C… grey, H … cyan, O … red, Se … yellow, Zn … purple, total … dashed).*

### 2.5 Conduction band of $Zn_2$(TTFTB) and model TTF stacks

Considering the conduction band of $Zn_2$(TTFTB) (without water; mFigure S7) we find that it has a significantly smaller band width than the valence band. This has already been discussed in the main manuscript. Additionally, we can observe that bands along directions perpendicular to the stacking direction (KΓ and AL) exhibit small dispersions of around 20 meV. This means that unlike for the valence band, for the conduction there is a small coupling between neighboring TTF stacks. This coupling is potentially mediated by Zn and O p-states, as can be rationalized by these atoms'/orbitals' contributions to the conduction band.

Considering that not solely states arising from the central TTF core contribute to the conduction band it is not surprising that a model stack consisting only of TTF molecules cannot reproduce the conduction bands of the actual MOF (see dashed line in Figure S7).

Interestingly, for the system with water (after relaxing the atomic positions), we find that the dispersion along KΓ and AL is significantly reduced (see Figure S8). This indicates a weaker coupling between neighboring TTF stacks for the conduction band in the presence of water.

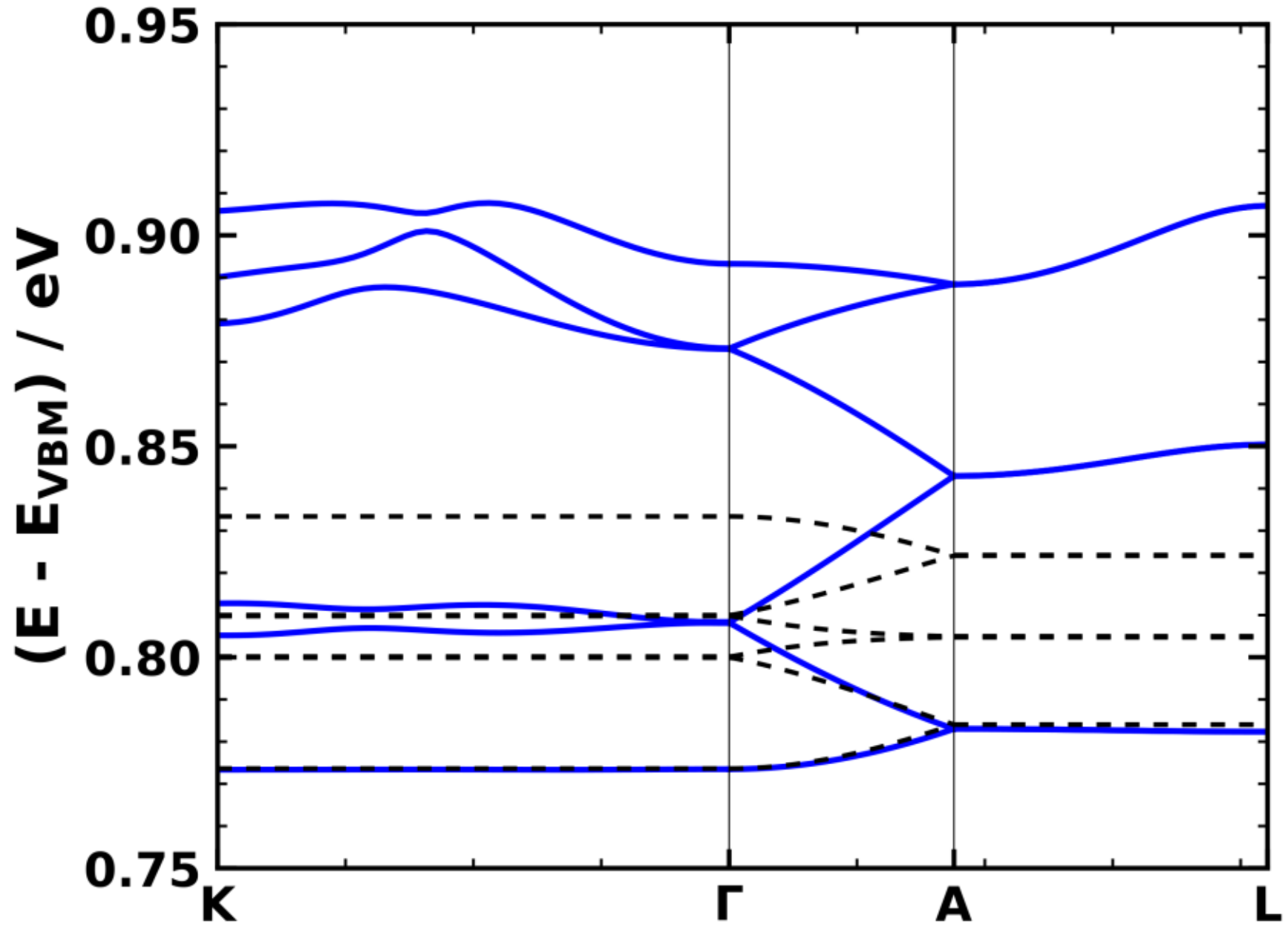

***Figure S7.*** *Electronic band structure of the conduction band for $Zn_2$(TTFTB) in blue and the corresponding TTF model stack as the black dashed line. One can see that bands along reciprocal space directions perpendicular to the TTF stacks show a small dispersion (~20 meV), which means that for the conduction band there is a small coupling between neighboring stacks.*

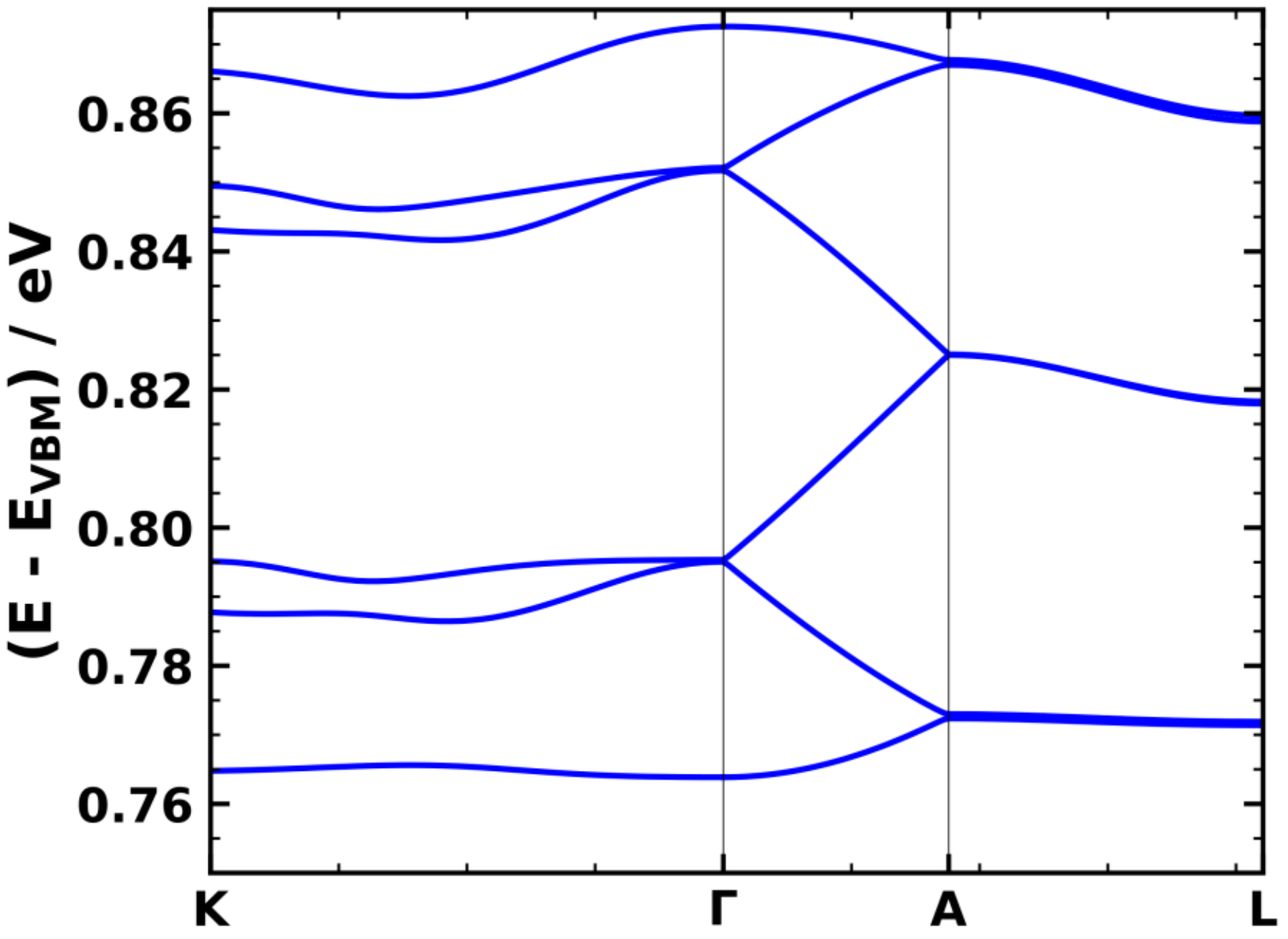

***Figure S8.*** *Electronic band structure of the conduction band for $Zn_2$(TTFTB) with water coordinated to the Zn atoms. One can observe that in comparison to the system without water, the dispersion along directions perependicular to the TTF stacks is reduced significantly.*

**2.6 Electronic structure of a stack consisting of the central TTF core of the TTFTB linkers plus the phenylene rings**

From the projected density of states in the main manuscript one can already conclude that the phenyl rings show a non-negligible contribution to the valence band of $Zn_2$(TTFTB). To show this in a more explicit way we, extracted a stack of TTFTB molecules from the MOF and replaced the carboxyl groups with H (see structure in Figure S9). After relaxing the atomic positions of these H atoms we calculated the electronic band structure of this model system. Comparing the valence bands of the model system and of the full $Zn_2$(TTFTB) MOF in Figure S9 reveals an excellent agreement between the band structures of the two systems.

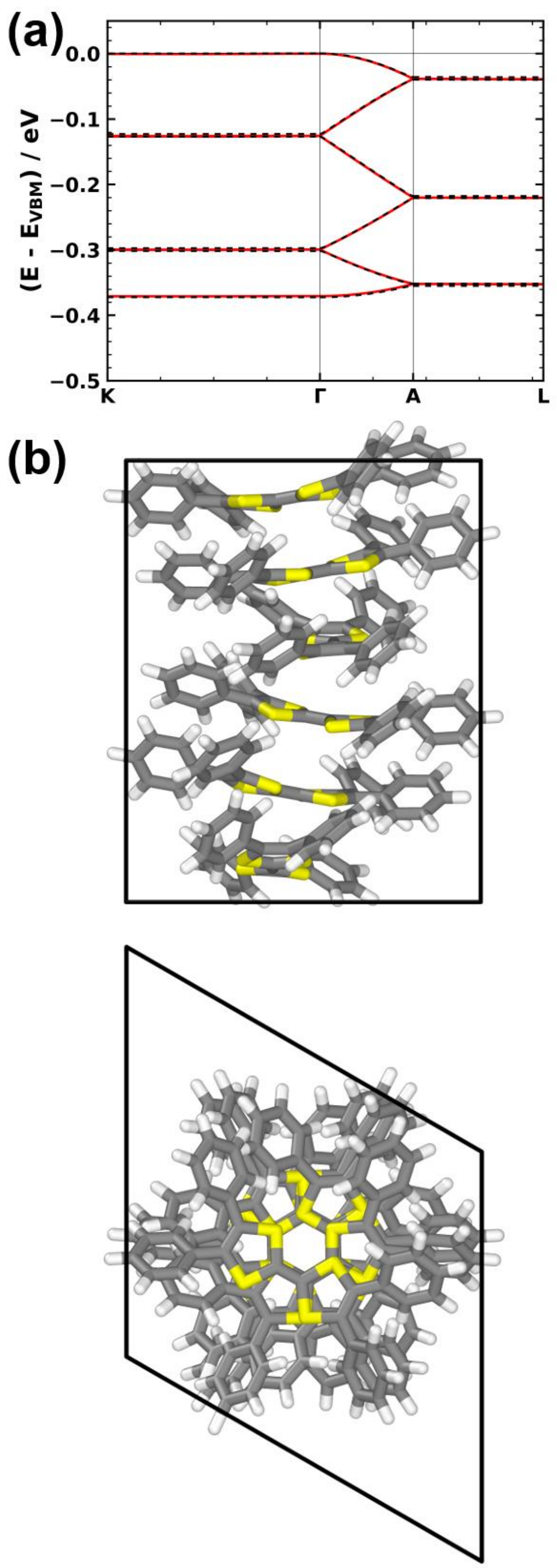

***Figure S9.*** *(a) Electronic band structure of $Zn_2$(TTFTB) in red and of the isolated and saturated TTFTB stack (black). (b) In the lower panel the unit cell of the TTFTB stack is shown.*

## 2.7 Electronic band structures of the considered TTF stacks

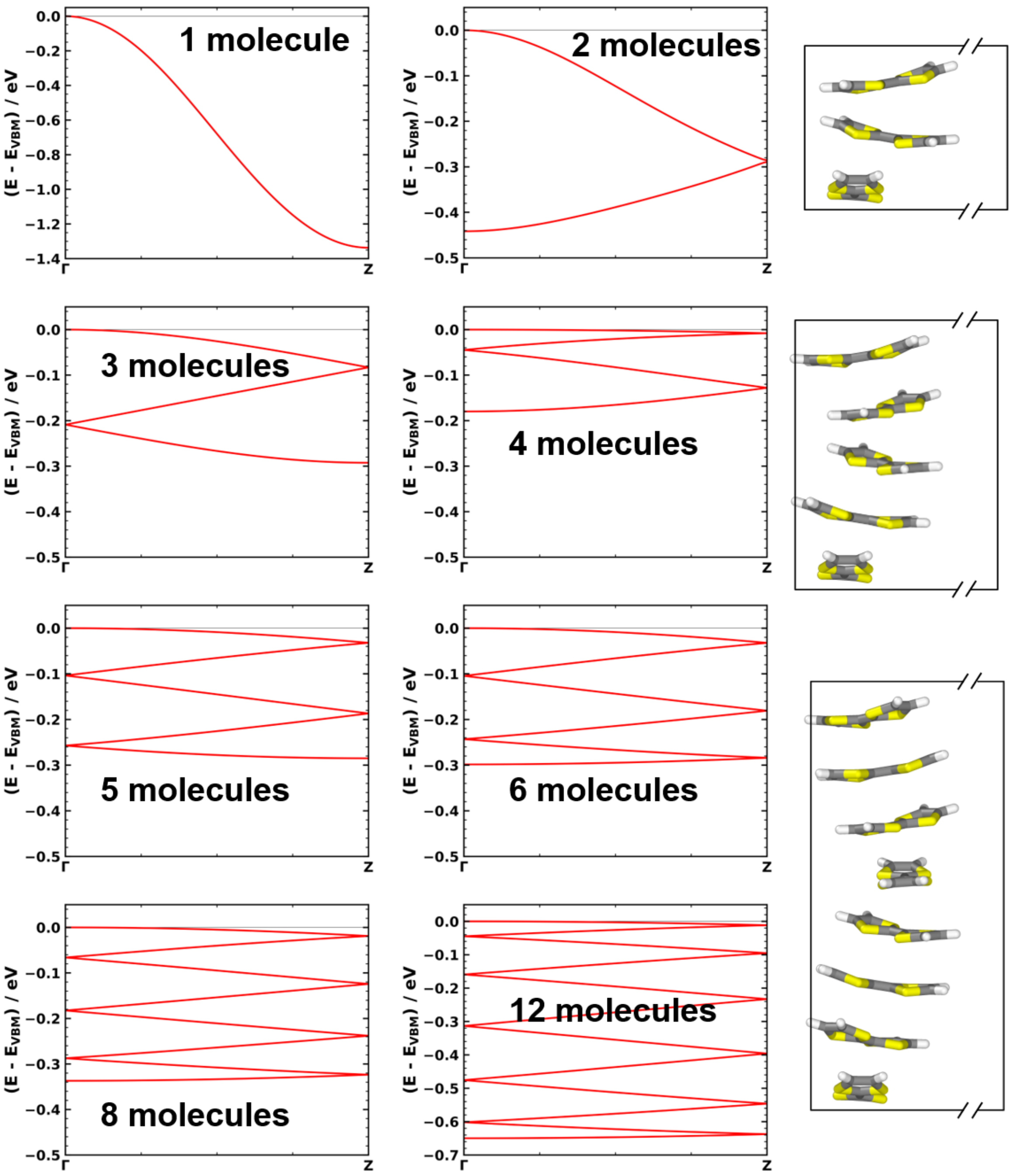

***Figure S10.*** *Electronic band structures of the TTF model stacks with 1, 2, 3, 4, 5, 6, 8, 12 repeat units. In the right panel unit cells for systems with 3, 5, and 8 repeat units are shown as examples.*

## 2.8 Electronic band structures deviating from the simple tight-binding picture

For certain systems we found that the electronic band structure of the valence band deviates from the shape one would expect for a simple 1D tight-binding model (see section 1.2 of SM). For relaxed with 4, boat and planar molecular geometries with 5 repeat units we find that the band maximum is slightly off Γ, but with an energy difference significantly smaller than 25 meV ($k_BT$ at room temperature). In these systems the n-fold backfolded valence band splits into two electronic bands. For determining the band width we, thus, consider the entire energy range covered by these bands. For planar TTF with 3 repeat units we see that the band has picked up a contribution with a higher frequency, meaning that electronic couplings beyond the nearest neighbor become relevant.[60]

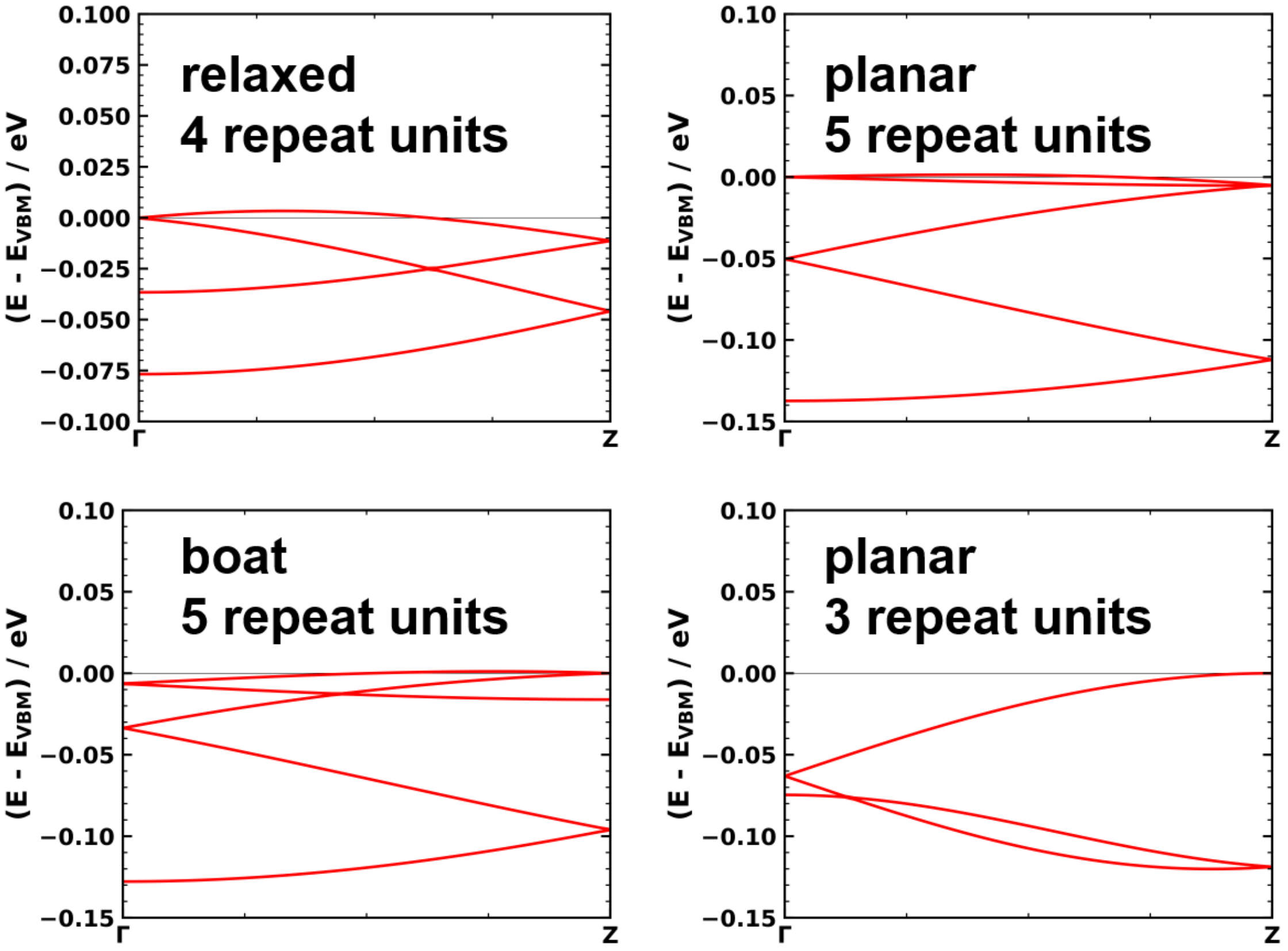

***Figure S11.*** *Electronic band structure of the stacks marked by black frames in Figure 3 of the main manuscript. For these systems we find that the electronic band structures deviate from the shape one would expect from a simple 1D tight-binding model with a single TTF molecule as "electronic" repeat unit.*

## 2.9 Total energies of the optimized TTF stacks

Considering the total energies per TTF repeat unit of the relaxed TTF model stacks (see Figure S12a) we find that except for the cofacial system (1 repeat unit) and the the 2 repeat unit system all stacks exhibit energies within less than 35 meV. Considering the planar TTF model system shown in Figure S12b one can find a correlation between the total energy and the transfer integral (band width), especially for large rotation angles. Such a behavior is, in fact, not unexpected considering the role of exchange interactions as described in [33]. One can, however, also observe that the vdW interactions between the TTF molecules play an important role in determining the energetic stability of certain arrangements (TTF stacks). For the fully optimized structures, such a correlation is less pronounced, which is a consequence of different distortions of the molecules at different rotation angles significantly changing the distances between atoms in neighboring molecules.

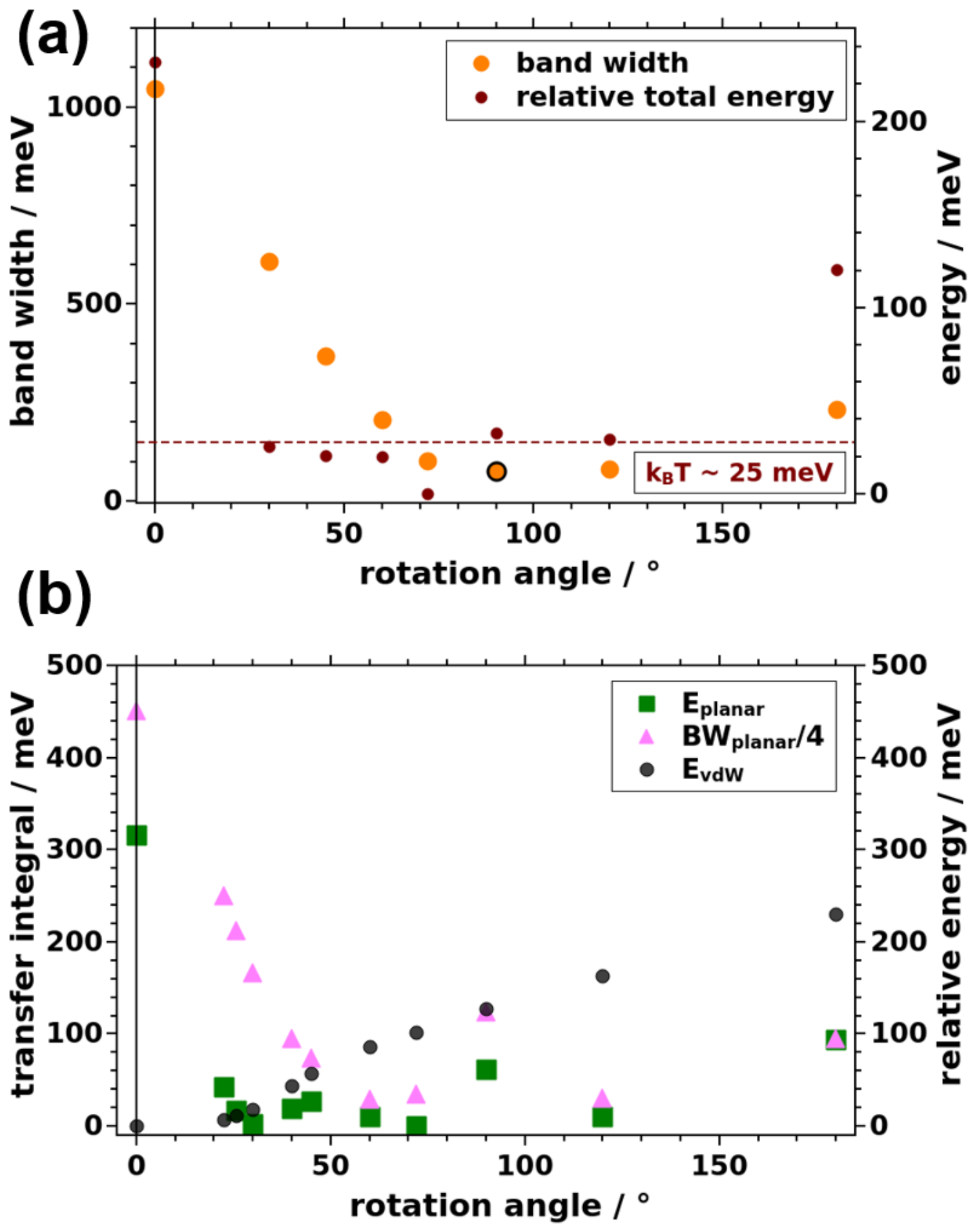

***Figure S12.*** *Evolution of the band width (transfer integral) and the total energy per TTF molecule for theTTF model stacks. (a) Band width and relative total energy for the relaxed*

*TTF model stack. (b) Transfer integral as (1/4)* valence band width, relative total energy, and vdW energy for the planar TTF model stack. For the planar model stack additional data points for 9, 14, and 16 repeat units were added. This was done to clarify the evolution of the relative total energy for angles between 20° and 50°. The energies are aligned to the global minimum, this means that positive energies result in less stable arrangements.*

## 2.10 Transfer Integral for a planar TTF dimer with a centered rotation axis

To investigate the influence of the position of the rotation axis on the evolution of the transfer integrals we considered a planar TTF dimer with the rotation axis placed in the center of the molecules. Varying the relative rotation angles of the two monomers with respect to this rotation axis results in the evolution shown in Figure S13a. One can see that for this system the orbitals never change their order, consequently also the transfer integral always exhibits the same sign. The respective orbitals for the extrema of the transfer integral are shown in Figure S13b.

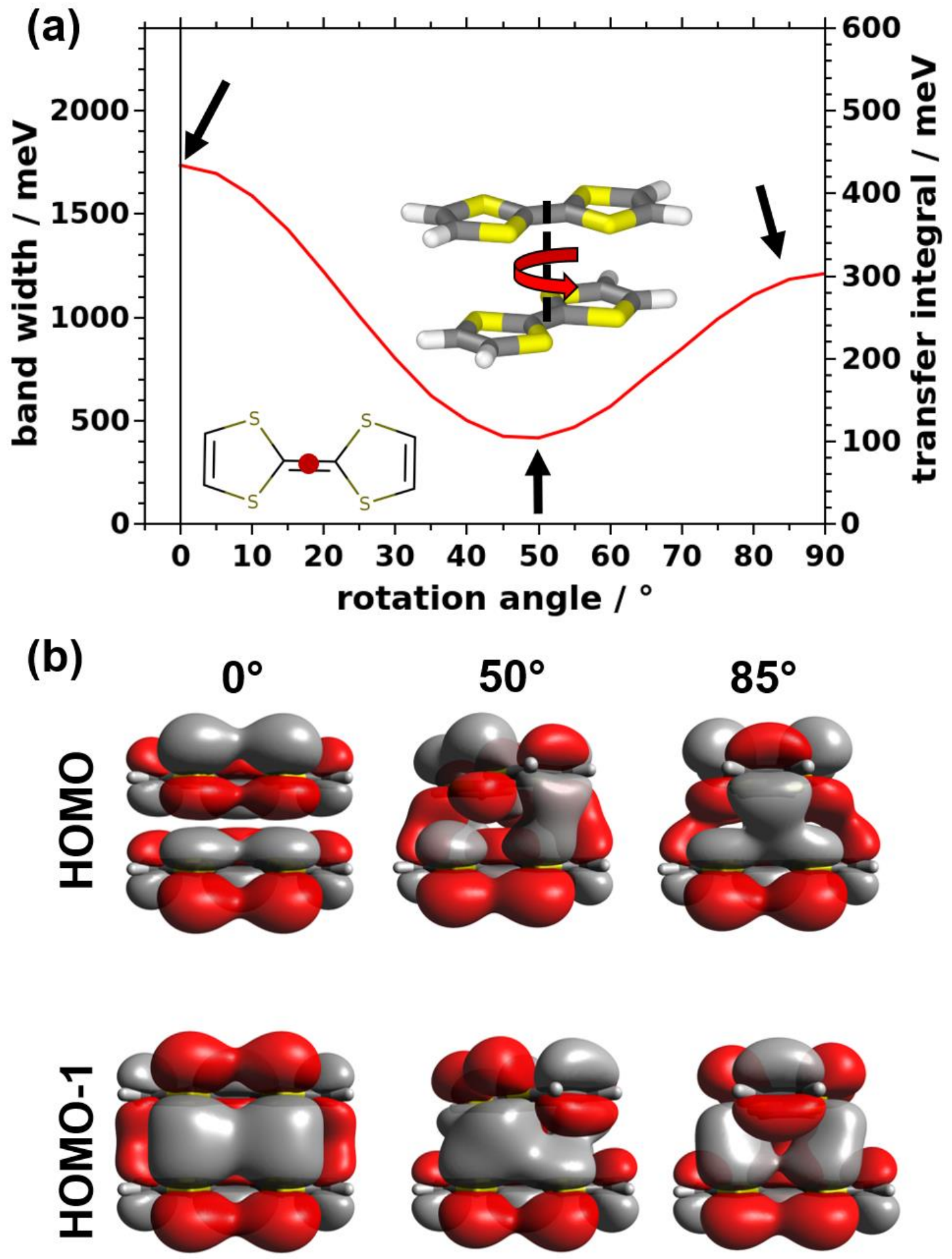

*Figure S13.* Electronic structure of Planar planar TTF dimers with the rotation axis placed in the center of the TTF molecules. Panel (a) shows the evolution of the band-width and the transfer integral as a function of rotation angle. Panel (b) shows the molecular orbitals (HOMO and HOMO-1) for rotation angles corresponding to local the extrema of the band-width (marked with black arrows).

## 3. Data for additional systems considered in the current manuscript: $Cd_2$(TTFTB), $Zn_2$(TSFTB)

### 3.1 Electronic structure of $Cd_2$(TTFTB)

As an additional MOF system considered here is $Cd_2$(TTFTB) (structure from ref [49]), which is isostructural to $Zn_2$(TTFTB). For this system we relaxed the atomic positions within the reported unit cell[49] and then calculated the electronic band structure for this relaxed geometry. The resulting species projected density of states and the valence band structure are shown in Figure S13.

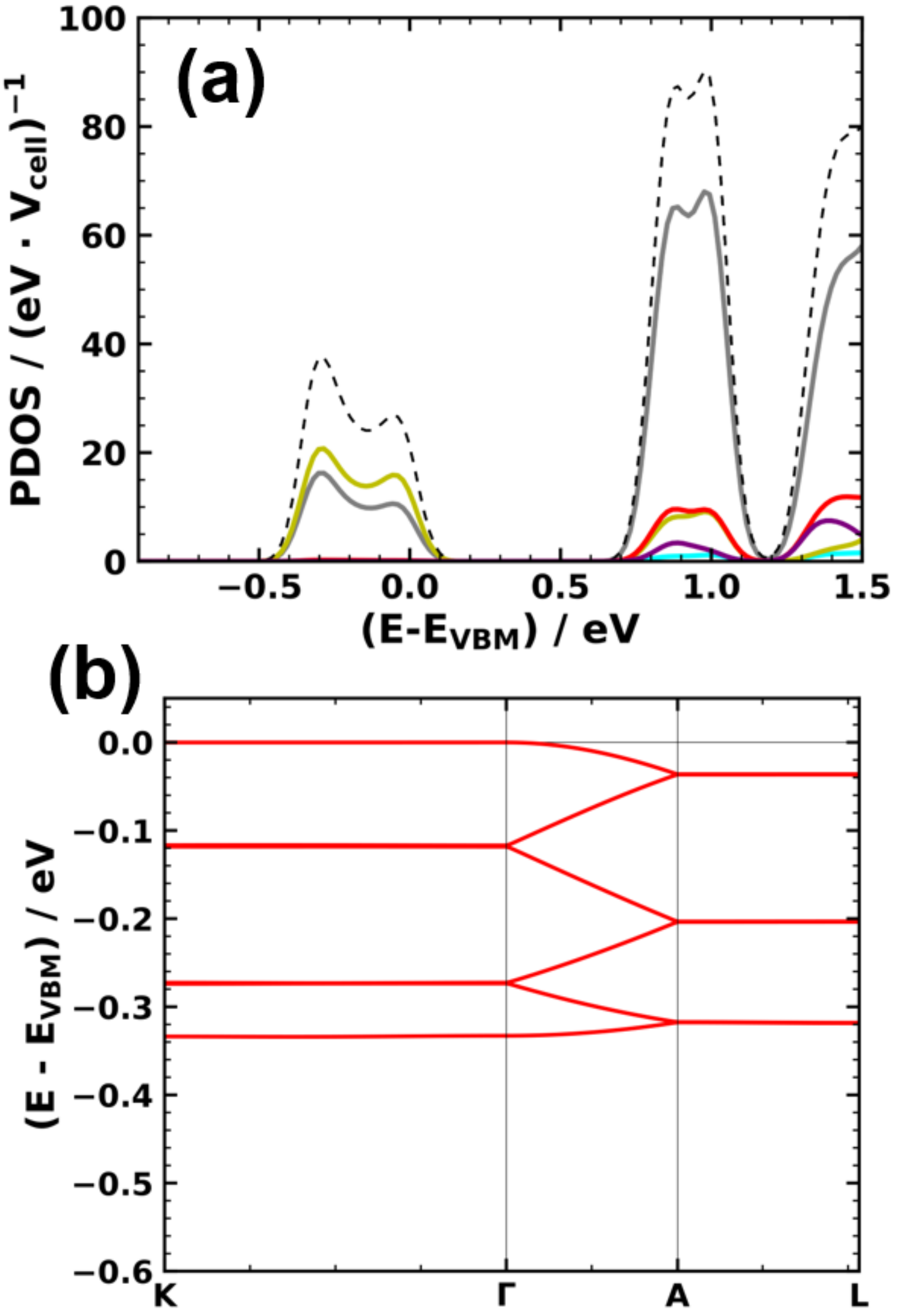

***Figure S14.*** *Electronic structure of the $Cd_2$(TTFTB) MOF. (a) species projected density of states (C… grey, H … cyan, O … red, Se … yellow, Cd … purple) and (b) valence band aligned to its maximum.*

### 3.2 Electronic structure of $Zn_2$(TSFTB)

Also the electronic structure (PDOS and electronic band structure) of a MOF with Tetraselenafulvalene ($C_6H_4Se_4$) replacing TTF has been calculated. The structure of this system was obtained by taking the structure of $Zn_2$(TTFTB) and replacing S with Se, i.e. replacing TTF with TSF. As no experimental cell for this system exists and as we expect the larger p-orbitals of Se to cause an increase of the stacking distance, we relaxed the atomic positions of the starting geometry as well as the cell vectors. For comparison we also calculated the electronic for the system when relaxing the atomic positions while keeping the unit cell fixed. The obtained species projected density of states and the structure of the valence band for the system with the relaxed unit cell are shown in Figure S15. The data for the system with relaxed atomic positions but within the unit cell of $Zn_2$(TTFTB) are shown in Figure S16. For the fully relaxed system we obtain 678 meV for the valence band width and an effective mass of 0.98 $m_e$. For the system in the $Zn_2$(TTFTB) unit cell we get 641 meV and 1.06 $m_e$, so there is hardly any difference between the relevant quantities of these systems.

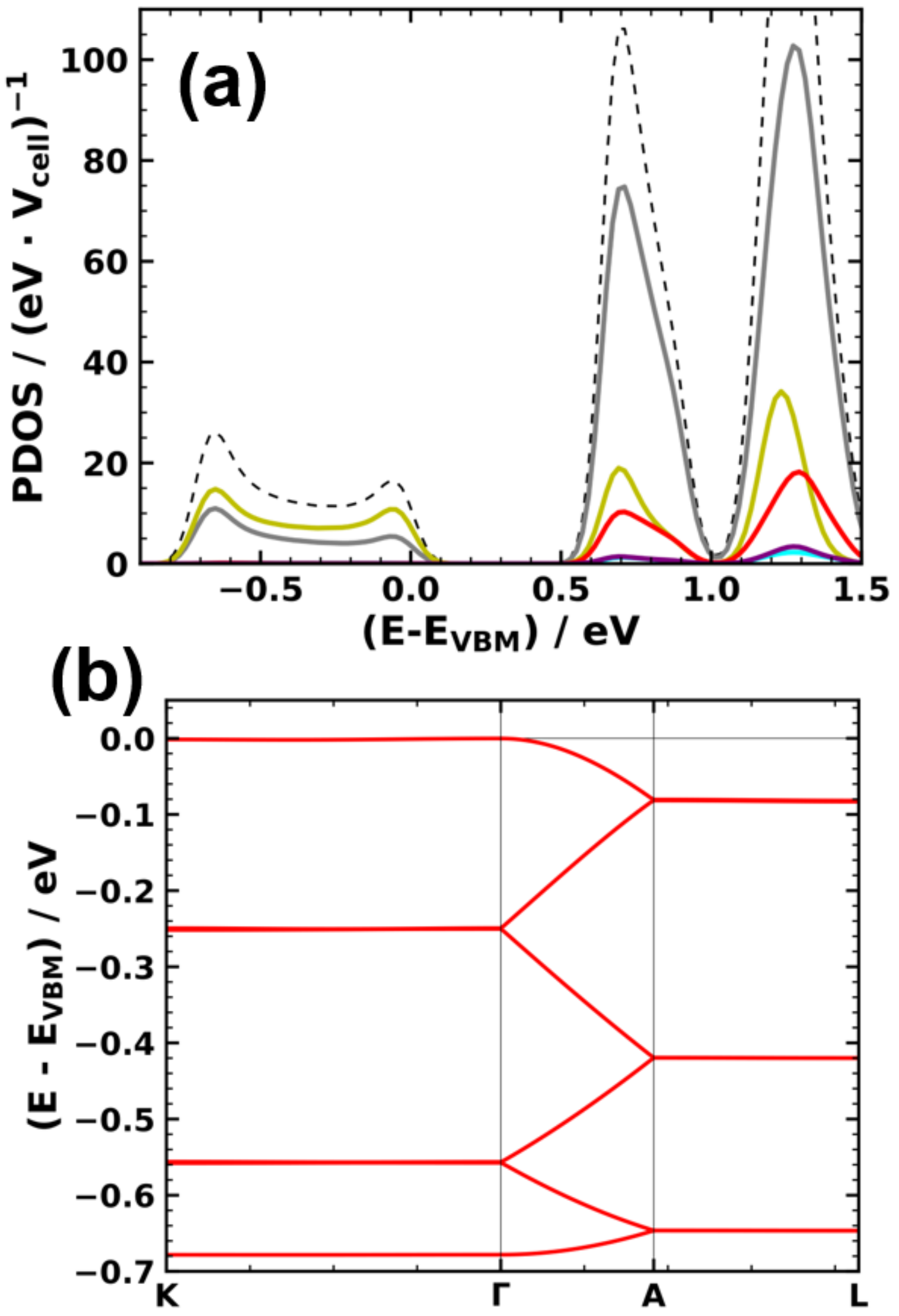

***Figure S15.*** *Electronic structure of the $Zn_2$(TSFTB) MOF for the relaxed unit cell. (a) Projected density of states (C… grey, H … cyan, O … red, Se … yellow, Zn … purple, total … black dashed) and (b) electronic band structure of the valence band aligned to its maximum.*

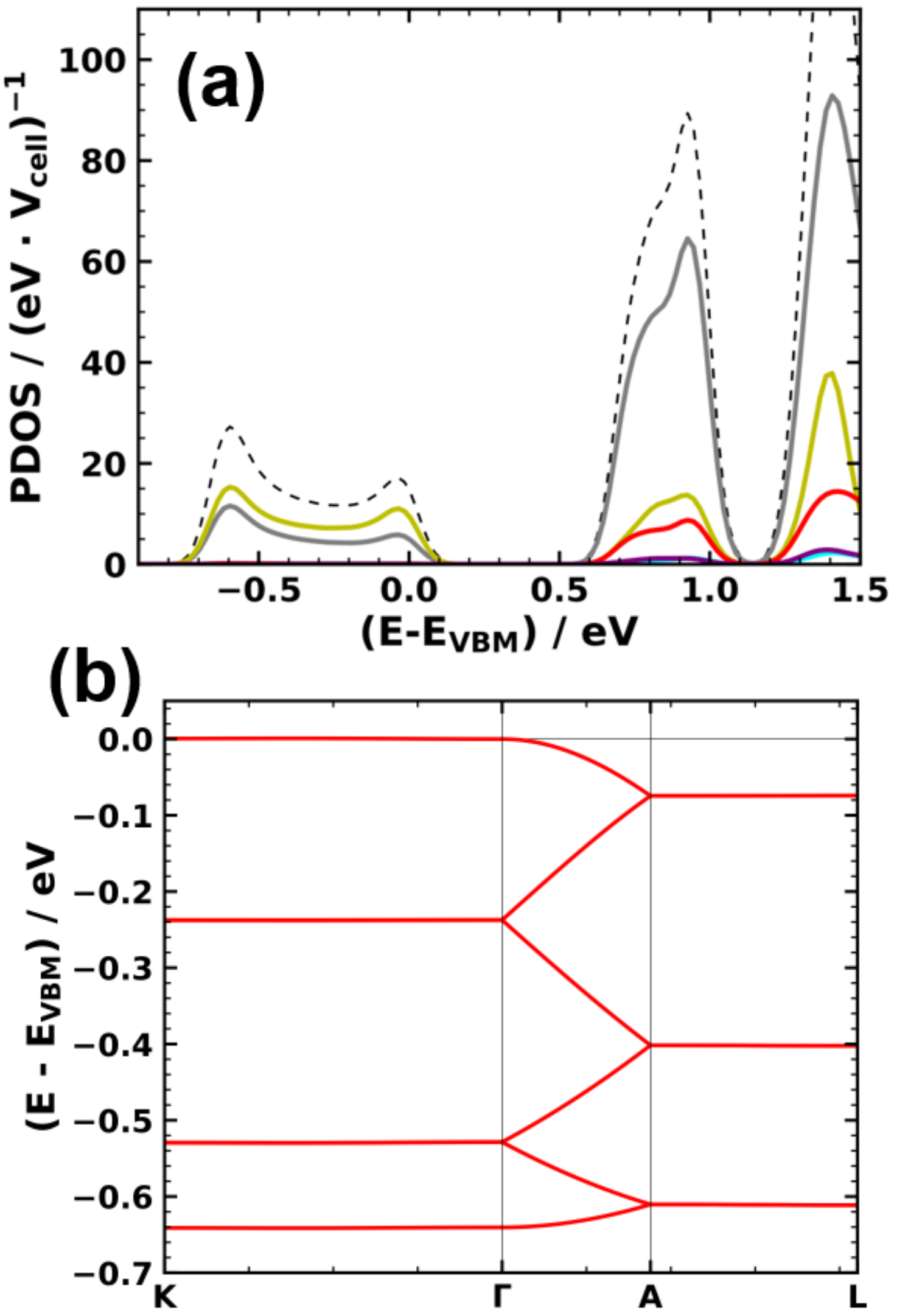

***Figure S16.*** *Electronic structure of $Zn_2$(TSFTB) MOF in the unit cell of $Zn_2$(TTFTB). (a) Projected density of states (C… grey, H … cyan, O … red, Se … yellow, Zn … purple, total … black dashed) and (b) electronic band structure of the valence band aligned to its maximum.*

# 4. Additional data for defects within the MOF and the model systems

## 4.1 Electronic structure of a TTF model systems with a displaced rotation axis

To show the influence of other potential defects we considered also shifts of the center of rotation for one of the TTF molecules. The resulting band structures are shown in Figure S17. Again one can see that upon introducing of such defects a gaps open at the BZ boundary and at the Γ point. These gaps and the resulting changes of the band dispersion y lead to an increase of the effective mass (Table S2), i.e a decrease of the transfer integral – similar to the data presented in the main manuscript.

***Table S2.*** *Effective mass depending on the offset of the rotation center of one of the TTF molecules in the model stack.*

| **Shifts (Δx, Δy) / Å** | **Effective mass m* / me** |
|---|---|
| (0.0, 0.0) | 2.48 |
| (0.0, 0.1) | 2.55 |
| (-0.1, 0.1) | 2.57 |
| (-0.25, 0.25) | 3.28 |
| (-0.5, 0.5) | 4.70 |
| (-1.25, 1.40) | 12.62 |

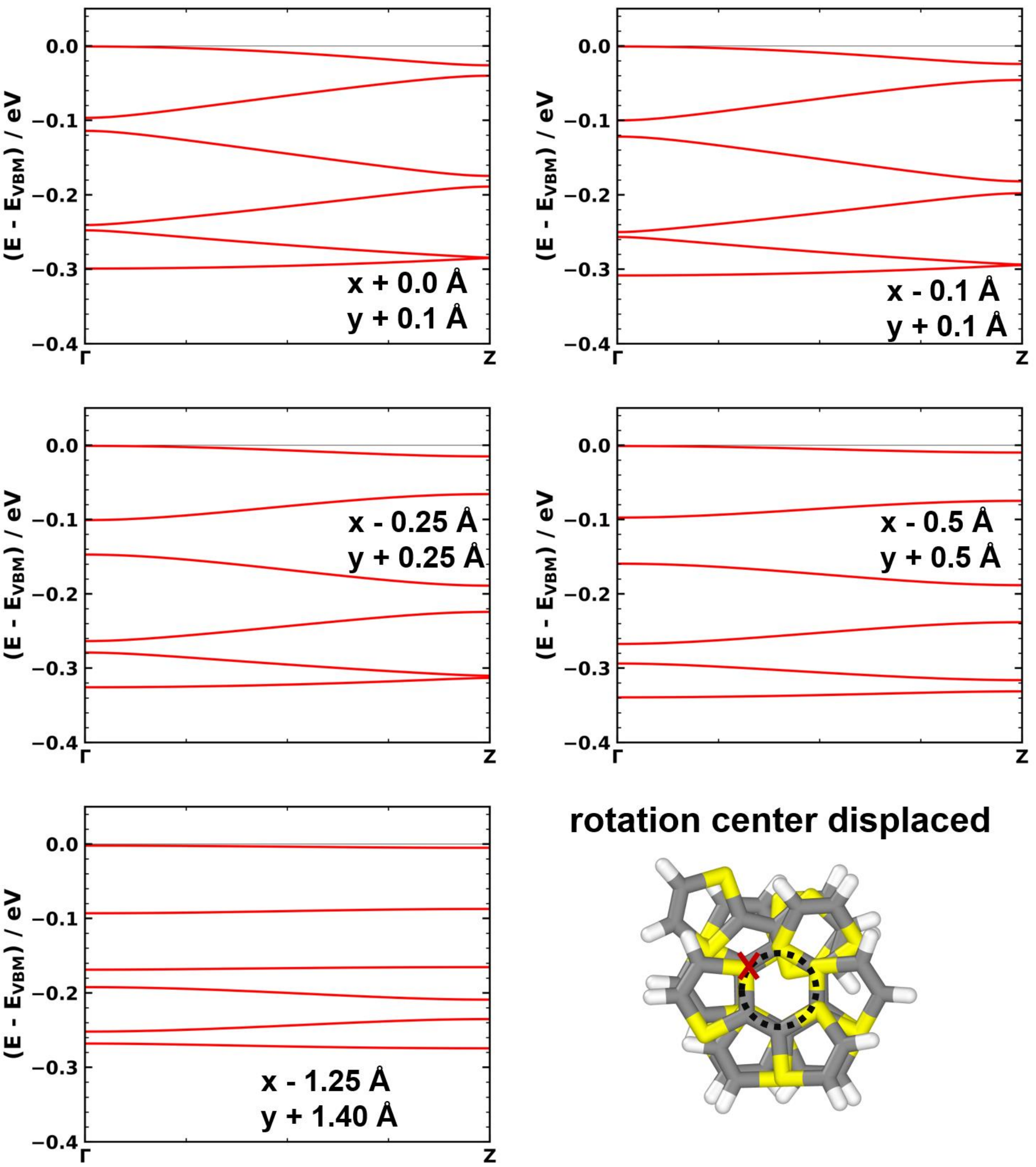

***Figure S17.*** *Electronic band structures for TTF model stacks where the rotation axis for one of the molecules is displaced from the equilibrium position.*

## 4.2 Electronic band structures for the dimerization defect data

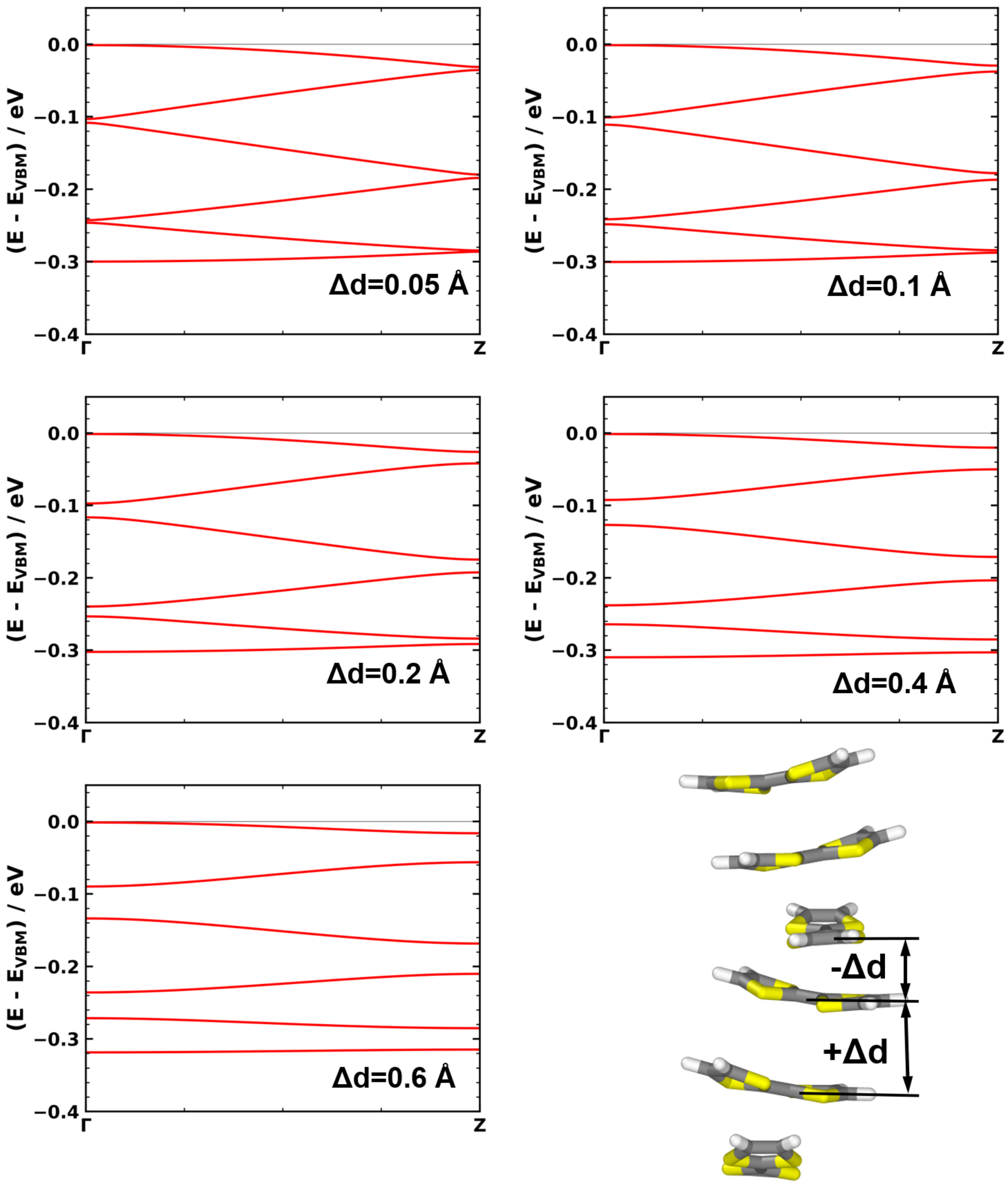

***Figure S18.*** *Electronic band structure of the valence band for the 6 repeat units TTF model stack with dimerization defects Δd.*

## 4.3 Electronic band structures for the "displaced molecule" defect data

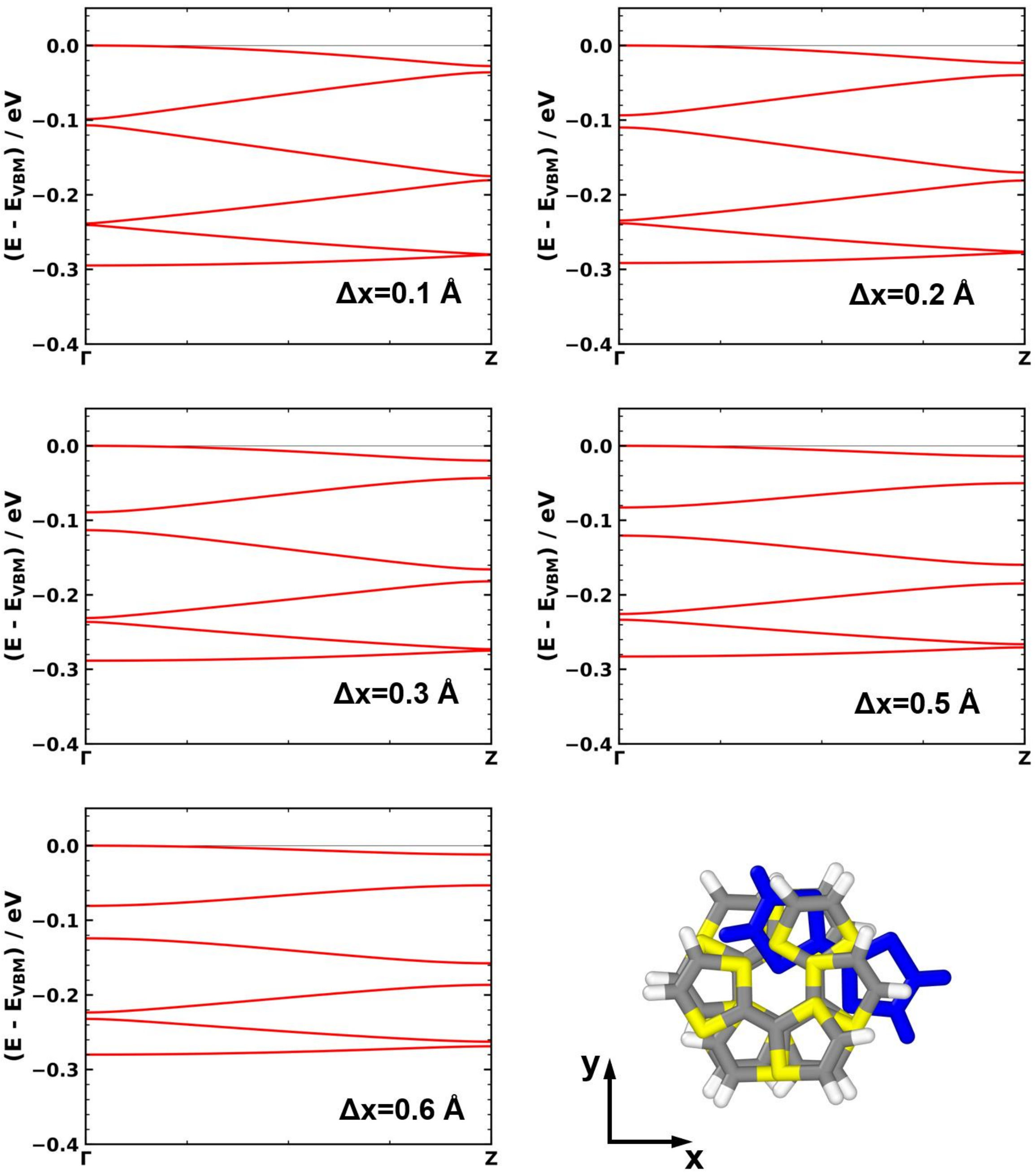

***Figure S19.*** *Electronic band structure of the valence band for the 6 repeat units TTF model stack with "displaced molecule" defects Δx.*